\documentclass[fleqn,10pt]{wlscirep}
\usepackage[utf8]{inputenc}
\usepackage[T1]{fontenc}
\usepackage{lineno}

\usepackage[version=4]{mhchem}
\usepackage{siunitx}
\usepackage{xspace}
\usepackage{multirow}
\usepackage{amsmath}
\usepackage{bm}

\newcommand{\x}{\ensuremath{x}\xspace}
\newcommand{\arcsinh}{\ensuremath{\mathrm{arcsinh}}\xspace}

\newcommand{\Tc}{\ensuremath{T_\sub{c}}\xspace}
\newcommand{\Tcry}{\ensuremath{T_\sub{cry}}\xspace}
\newcommand{\scmp}{\ensuremath{\mathrm{3DSC_{MP}}}\xspace}
\newcommand{\scicsd}{\ensuremath{\mathrm{3DSC_{ICSD}}}\xspace}
\newcommand{\scdataset}{\ensuremath{\mathrm{3DSC}}\xspace}

\newcommand{\totreldiff}{\ensuremath{\Delta_\sub{totrel}}\xspace}
\newcommand{\fsym}{\ensuremath{F_\mathrm{sym}}\xspace}
\newcommand{\maxtotreldiff}{\ensuremath{\totreldiff ^\sub{max}}\xspace}
\newcommand{\ehull}{\ensuremath{E_\mathrm{hull}}\xspace}
\newcommand{\efermi}{\ensuremath{E_\mathrm{F}}\xspace}

\newcommand{\explicittcry}{\ensuremath{\mathrm{\Tcry ^\sub{explicit}}}}
\newcommand{\sub}[1]{\mathrm{#1}}

\newcommand{\ybco}{\ce{YBa2Cu3O7}\xspace}
\DeclareRobustCommand{\kelvin}[1]{\ensuremath{\SI{#1}{\kelvin}}\xspace}
\addto\extrasenglish{

}

\title{3DSC - A New Dataset of Superconductors Including Crystal Structures}

\author[1,2,5]{Timo Sommer}
\author[2]{Roland Willa}
\author[2,3]{Jörg Schmalian}
\author[1,4,*]{Pascal Friederich}
\affil[1]{Institute of Theoretical Informatics, Karlsruhe Institute of Technology, Am Fasanengarten 5, 76131 Karlsruhe, Germany}
\affil[2]{Institute for Theory of Condensed Matter, Karlsruhe Institute of Technology, Wolfgang-Gaede-Str. 1, 76131 Karlsruhe, Germany}
\affil[3]{Institute for Quantum Materials and Technologies, Karlsruhe Institute of Technology, Hermann-von-Helmholtz-Platz 1, 76344 Eggenstein-Leopoldshafen, Germany}
\affil[4]{Institute of Nanotechnology, Karlsruhe Institute of Technology, Hermann-von-Helmholtz-Platz 1, 76344 Eggenstein-Leopoldshafen, Germany}
\affil[5]{School of Chemistry, Trinity College Dublin, College Green, Dublin 2, Ireland}

\affil[*]{corresponding author(s): Pascal Friederich (pascal.friederich@kit.edu)}

\begin{abstract}
Data-driven methods, in particular machine learning, can help to speed up the discovery of new materials by finding hidden patterns in existing data and using them to identify promising candidate materials. In the case of superconductors, which are a highly interesting but also a complex class of materials with many relevant applications, the use of data science tools is to date slowed down by a lack of accessible data. In this work, we present a new and publicly available superconductivity dataset ('3DSC'), featuring the critical temperature \Tc of superconducting materials additionally to tested non-superconductors. In contrast to existing databases such as the SuperCon database which contains information on the chemical composition, the 3DSC is augmented by the approximate three-dimensional crystal structure of each material. We perform a statistical analysis and machine learning experiments to show that access to this structural information improves the prediction of the critical temperature \Tc of materials. Furthermore, we see the 3DSC not as a finished dataset, but we provide ideas and directions for further research to improve the 3DSC in multiple ways. We are confident that this database will be useful in applying state-of-the-art machine learning methods to eventually find new superconductors.
\end{abstract}

\begin{document}

\flushbottom
\maketitle

\thispagestyle{empty}

\section{Introduction}

Superconductors are materials in which the electrical resistance is zero when the temperature drops below a critical temperature \Tc. Furthermore, superconductors are perfect diamagnets that expel magnetic fields via the Meissner effect. These properties make superconductors very useful for many high-power applications such as efficient electric power conversion, lossless power transmission, and ultra-strong magnets, as well as high-sensitivity sensor materials e.g.\ superconducting quantum interference devices and photon detectors \cite{yaoSuperconductingMaterialsChallenges2021,eleyChallengesTransformativeOpportunities2021}. The discovery of new superconducting materials with optimized properties will enable e.g.\ the use of cheaper coolants due to increased critical temperatures, stronger magnets due to improved magnetic properties, and simpler production of superconducting wires due to improved mechanical properties.

The critical temperature \Tc can be very sensitive to small changes in the crystal structure, for example to changes in the interatomic distances via mechanical pressure or chemical pressure\cite{horHighpressureStudyNew1987}, i.e. the deformation of the lattice by replacing one atom with another element with same valency but different size.
Despite the success of understanding the mechanism behind superconductivity within a microscopic theory, such as  the theory by Bardeen, Cooper and Schrieffer \cite{bardeenMicroscopicTheorySuperconductivity1957} and strong-coupling generalizations thereof, we are to date unable to faithfully predict the critical temperature \Tc of new materials. This is largely caused by the dependence of \Tc on subtle details of atomic arrangements in the crystal structure.
Input parameters of the microscopic, low-energy theories, such as the electronic density of states at the Fermi level, the phonon spectrum, and the electron-lattice coupling, are not easily related to the chemical formula. 
Thus, when predicting the critical temperature \Tc with machine learning, a first step to narrow this gap is to have access not only to the chemical composition of the material, but also to the exact 3D structure of the crystal.

Machine learning has been widely used for the prediction of materials properties. Saal et al.\cite{saalMachineLearningMaterials2020} collected and reviewed a big number of machine learning generated predictions which have been confirmed experimentally afterwards in applications ranging from organic LEDs over new binary and ternary crystal structures, perovskites, metallic glasses and metal-organic-frameworks to superhard materials. Furthermore, machine learning was used to predict the critical temperature \Tc of superconductors using the SuperCon database\cite{SuperCon2020} as training data, which is the largest and most commonly used dataset of superconductors. Unfortunately, the SuperCon database is unavailable as of December 2021. However, previous papers have made parts of the preprocessed data available for further research\cite{vstanev1Supercon2021, hamidiehDataDrivenStatisticalModel2018}. 

There have been many attempts to predict the critical temperature \Tc of a material using the SuperCon database. Hamidieh\cite{hamidiehDataDrivenStatisticalModel2018} used a gradient boosting model (XGB\cite{chenXGBoostScalableTree2016}) trained on MAGPIE features\cite{wardGeneralpurposeMachineLearning2016b} to predict \Tc. Aketi et al.\cite{aketiREGRESSIONSUPERCONDUCTINGCRITICAL2019} used gradient boosted decision trees, Matsumoto et al.\cite{matsumotoAccelerationSearchMethod2019} used random forests and Le et al.\cite{leCriticalTemperaturePrediction2020} used Bayesian neural networks on very similar features, while Gaikwad et al.\cite{gaikwadFeaturelessApproachPredicting2020} compare multiple machine learning models. Konno et al.\cite{konnoDeepLearningModel2021} and Zeng et al.\cite{zengAtomTableConvolutional2019} used a convolutional neural network (CNN) and represented the chemical formula as elements on a grid. Li et al.\cite{liCriticalTemperaturePrediction2020} used a hybrid neural network consisting of a CNN and a recurrent neural network (RNN) which is trained on Atom2Vec features\cite{zhouAtom2VecLearningAtoms2018}. Dan et al.\cite{danComputationalPredictionCritical2020} use a convolutional gradient boosting decision tree (ConvGBDT). Sizochenko et al.\cite{sizochenkoPredictiveModelingCritical2021} found that an often used subset of the SuperCon contained a lot of duplicate entries and repeated their analysis with the cleaned dataset. Meredig et al.\cite{meredigCanMachineLearning2018} showed that random splits for the cross validation give overly confident model evaluations. Roter et al.\cite{roterPredictingNewSuperconductors2020} trained a bagged tree model on the chemical composition and argued that physical features such as the Fermi energy would be helpful for increasing the performance of their model if they were available for more materials. 

Data availability is the most important prerequisite for the development of (supervised) machine learning models for materials property prediction. In particular, informative and complete information on the materials is essential for the training of accurate machine learning models. All of the studies discussed above were based on representing materials only by their chemical composition, which is not a unique and complete representation of materials. Yet, most SuperCon entries contain only the chemical formula and critical temperature \Tc of each material. Structural data such as space group and crystal system are only sparsely recorded and the full three-dimensional crystal structure is never given. Therefore, all of the aforementioned predictions of the critical temperature using the SuperCon database were limited to representations of the chemical composition of each material.

One notable exception of using only chemical formulas to predict critical temperatures is the work of Stanev et al.\cite{stanevMachineLearningModeling2018}. They developed a superconductivity classifier based on matching the chemical compositions of materials in the SuperCon with the chemical compositions of materials in the AFLOW database and used tabular structural and electronic features such as the space group and the energy per atom as additional features. In this pioneering work 1500 materials could be matched, half of them being superconductors. Stanev et al.\ argued that structural information is helpful in predicting superconductivity, yet realized the issue of severely reducing the size of the dataset when doing this matching. As of today, the matched crystal structures were not published.

Recently, two more databases dealing with superconductors were presented in the literature. The \mbox{SuperMat}\cite{foppianoSuperMatConstructionLinked2021} database and the SC-CoMIcs\cite{yamaguchiSCCoMIcsSuperconductivityCorpus2020} database are corpora of manually annotated texts from papers about superconductors. The annotations consist of different entities such as chemical formula and critical temperature with which certain phrases in the texts have been labeled. These corpora can be used for tasks such as training a named entity recognition model such as SciBERT\cite{beltagySciBERTPretrainedLanguage2019} on automatically labeling new papers, which was demonstrated by Yamaguchi et al.\cite{yamaguchiSCCoMIcsSuperconductivityCorpus2020}. Foppiano et al.\cite{foppianoSuperMatConstructionLinked2021} also publicly provide their annotation procedure to encourage others to continue this work. So far, these annotated corpora are not publicly available. In the future, they might be useful to automatically extract information about superconductivity from literature.

Court et al.\cite{courtMagneticSuperconductingPhase2020} used the already trained Chem\-Data\-Extractor\cite{swainChemDataExtractorToolkitAutomated2016} to extract information of superconductors and magnetic materials from literature. They found approximately 20,400 superconductors and magnetic materials together with their chemical compositions and respective phase transition temperatures. The focus of the study was the prediction of the phase diagram of magnetic and superconducting materials. Furthermore, some of the entries were paired with crystal structures from the Crystallographic Open Database (COD). The authors provide a link to an interactive web app and the data, yet, the provided link is currently inactive. Another recently initiated superconductor database is the Superconducting Research Database\cite{SuperCon2020}. In this online database, superconductors can be submitted with their exact three-dimensional crystal structure and critical temperature \Tc. This database currently contains 14 superconductors which limits its usefulness for machine learning processes.

In this work, we extended the structure matching approach by Stanev et al.\cite{stanevMachineLearningModeling2018} to build a new database (called 3DSC) of experimentally tested superconducting and non-superconducting materials. This database is made publicly available. The 3DSC database features the critical temperature of superconductors as well as the approximated 3D crystal structure of each material. The core idea is to match materials in the SuperCon database with (modified) crystal structures of the Materials Project\cite{jainCommentaryMaterialsProject2013,MaterialsProject} and the Inorganic Crystal Structure Database\cite{bergerhoffInorganicCrystalStructure1983,ICSD} (ICSD). In addition to matching only exact chemical compositions (as in Stanev et al.\cite{stanevMachineLearningModeling2018}), we employ a systematic adaptation algorithm that approximates the three-dimensional crystal structures of materials without perfect match by artificial doping of similar crystal structures. For example, the crystal structure of the SuperCon entry \ce{CuLa_{1.95}Nd_{0.05}O_4} (which has no perfect match in the Materials Project database) is approximated by taking the 3D crystal structure of \ce{CuLa2O4} and partially replacing La with Nd at the respective crystal positions. This step is important to maximize the number of matched materials, since the SuperCon contains many entries with doped materials, which otherwise would mostly be discarded.

In this paper, we introduce and analyze two different 3DSC databases. Both are based on the SuperCon database, but one uses structures from the Materials Project (\scmp) and one uses structures from the ICSD (\scicsd). Using our matching and adaptation algorithm, we are able to match 5,759 (\scmp) and 9,150 (\scicsd) superconducting and non-superconducting materials from the SuperCon. We publicly provide the full \scmp dataset on
\url{https://github.com/aimat-lab/3DSC},
including the critical temperature \Tc and approximate three-dimensional crystal structures.
However, structures from the ICSD must not be (re-)published.  Therefore, we refrain from publishing the \scicsd. The subset of the \scicsd that we provide under this link only contains the ICSD IDs necessary for reproducing the full dataset. The necessary structures can be downloaded with an ICSD license and artificially doped using our code in the aforementioned repository. However, despite not being able to publish this database, we have decided to present the \scicsd in this paper along with the \scmp, since it contains more structures and slightly different information than the \scmp. 

\begin{figure}[t]
\centering
\includegraphics[width=0.7\textwidth]{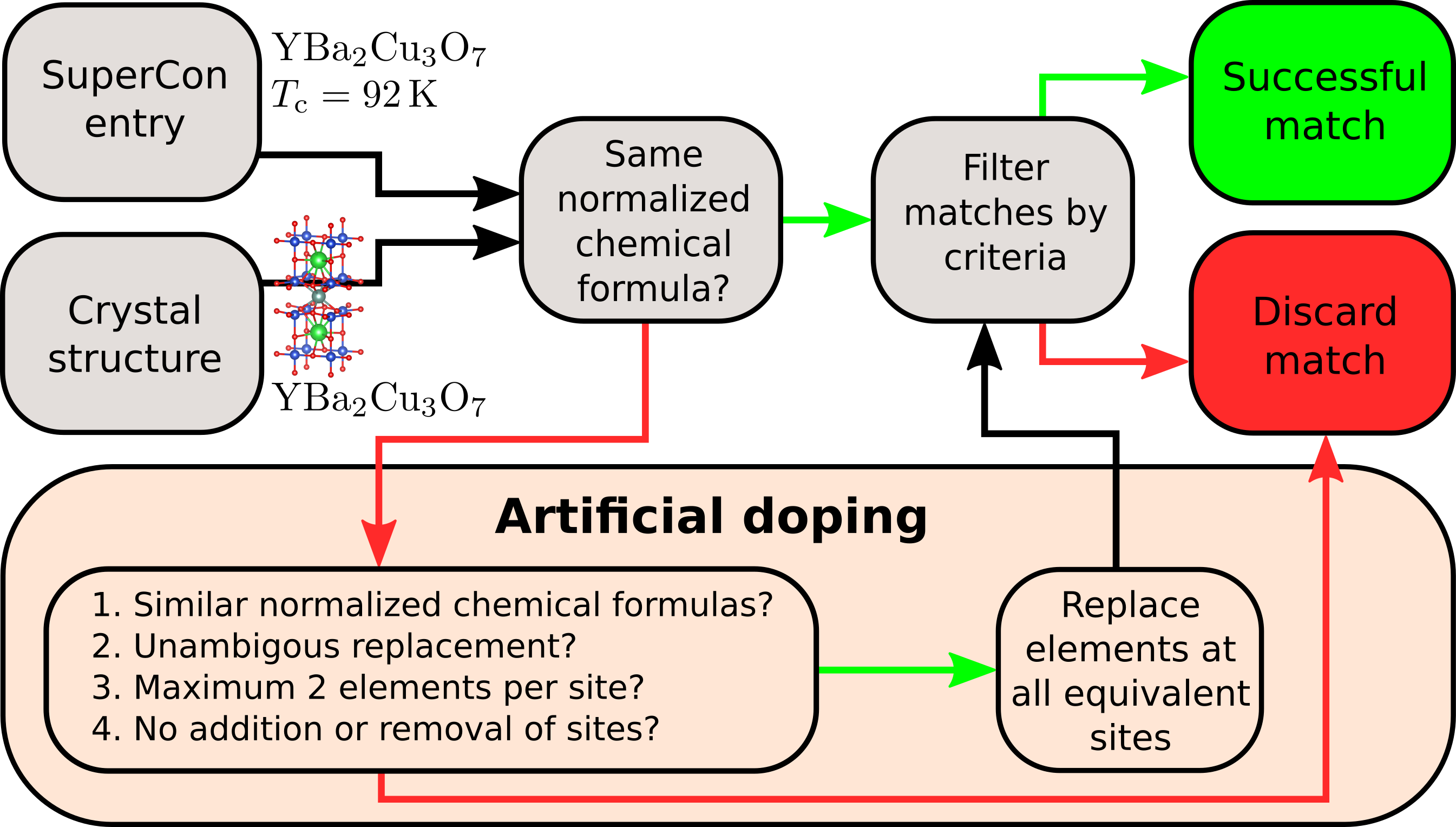}
\caption{A schematic view of the matching and adaptation algorithm.}
\label{fig:matching_algorithm}
\end{figure}

\section{Methods}

\subsection{Overview of 3DSC data generation}

In this section we describe our algorithm to match entries of the SuperCon database based on their chemical formula with 3D structures from crystal structure databases (see \autoref{fig:matching_algorithm}). We use and compare two different crystal structure databases, the Materials Project and the ICSD. We furthermore use the copy of the SuperCon database published by Stanev et al.\cite{stanevMachineLearningModeling2018}. All databases are cleaned as described in \autoref{sec:data}.

In the first step of the matching algorithm to build our 3DSC database, each SuperCon entry is paired with each crystal structure. If the chemical formula matches perfectly (after normalization as elaborated in \ref{si:normalizing_chemical_formulas}), the SuperCon entry is paired to the crystal structure and added to our database. If the chemical formula is close but not equal, we performed an artificial doping process to modify the crystal structure and to aim for a perfect match with the chemical formula of the SuperCon entry. This process is described in more detail in \autoref{sec:matching_algorithm} below.

Because this matching and adaptation algorithm generally matches multiple crystal structures to each SuperCon entry, we conclude the process by filtering the matches according to specific criteria and keeping only the most optimal matches. After applying this filter, we are left with the final 3DSC database which contains chemical formula, critical temperature \Tc, and the (in many cases approximate) crystal structure. In the case of the Materials Project database, the final \scmp dataset contains 5,759 SuperCon entries matched with 5,773 crystal structures. In the case of the ICSD, the final \scicsd dataset contains 9,150 SuperCon entries matched with 86,490 crystal structures. One reason for this high number of matched crystal structures is that the ICSD has a large number of entries with the same crystal structure at different crystal temperatures.

\subsection{Data and dataset cleaning}
\label{sec:data}

\paragraph{SuperCon:} 
The SuperCon database is the largest database of superconductors and has been used multiple times in the literature to predict the critical temperature of superconductors with machine learning methods. It contains approximately 33,000 materials that have been tested for superconductivity. Approximately 10,000 of the entries are duplicates of the same material with the same chemical formula. However, the SuperCon database features only the chemical formula of each material, whereas structural data such as space group and lattice-type is only sparsely recorded and the full crystal structure is never given. In this study, we use the already cleaned and published version of the SuperCon dataset published by Stanev et al.\cite{stanevMachineLearningModeling2018} which contains 16,400 different materials with approximately 4,000 non-superconductors. The dataset can be found on GitHub\cite{vstanev1Supercon2021} and includes the chemical formula and the critical temperature \Tc of each material.

We found that this dataset was not fully cleaned yet. We assume that in the original study chemical formulas were compared as strings, but sometimes the order of elements in the string was different even though it was the same material. For these 21 materials, we averaged the critical temperatures according to the same algorithm as used in Stanev et al., i.e.\ taking the mean and excluding the material if the standard deviation was greater than 5K. Because we are averaging over data points some of which are already averaged, the resulting average might not be the same as in the original dataset. Additionally, we found that some chemical formulas were invalid, e.g.\ in the formula \ce{Bi_{4.4}Sr_{3.6}Ca2Cu4OY} the Y was considered to represent yttrium, but it actually represented an unknown quantity of oxygen (the SuperCon database strictly follows a nomenclature, where each element has a count, even if the count is 1). Excluding these entries reduced the dataset by 128 entries. We also excluded 4 entries that had chemical formulas with more than 150 atoms because these are likely mistakes in the database. Finally we decided to exclude the few entries with the heavy elements americium (Am), curium (Cm) and polonium (Po) to reduce the number of entries with rarely occurring elements. After all cleaning steps, we are left with 15,758 SuperCon entries of which 3,854 are non-superconductors. The maximum critical temperature is \kelvin{143} for \ce{Ba2Ca_{1.98}Cu_{2.9}Hg_{0.66}Pb_{0.34}O_{8.4}}. Non-superconductors are encoded as having a critical temperature of \kelvin{0}. 

\paragraph{Crystal structure datasets:}
As sources for the crystal structures we used the Materials Project and the ICSD. The Materials Project contains approximately 139,000 DFT calculated structures and electronic features and is openly accessible. The ICSD contains approximately 243,000 mostly experimental structures and is accessible only with a license. The ICSD database was cleaned before further processing: 19,077 DFT calculated (rather than experimentally measured) entries were excluded to make the dataset more consistent. 18,147 entries were excluded because the chemical composition given by the ICSD and extracted from the crystal structure using the python package pymatgen\cite{ongPythonMaterialsGenomics2013} were inconsistent. Similarly, 3,195 entries were excluded because the space group given by the ICSD and the one recognized with pymatgen were inconsistent (likely due to numerical errors and thresholds). Finally, 3,132 more entries were excluded because of an invalid chemical formula. Even though not technically invalid, we decided to also exclude materials including deuterium and tritium. Because the ICSD also has the crystal temperature \Tcry recorded for most materials, entries without crystal temperature were assumed to be recorded at room temperature (\kelvin{293}).

The Materials Project was checked as well but no entries had to be excluded due to the aforementioned reasons. However, materials without recorded \ehull were excluded since this feature was needed for the matching and adaptation algorithm (see \autoref{sec:matching_algorithm}). Because the Materials Project structures had no crystal temperature given, the crystal temperature \Tcry was set to \kelvin{0} in order to have a consistent set of features for the machine learning models.

\subsection{Matching algorithm of SuperCon entries and 3D crystal structures}
\label{sec:matching_algorithm}

The following section describes the algorithm used for matching and adaptation (from now on referred to as \emph{artificial doping}) of SuperCon entries and crystal structures in more detail. The aim is to find one or multiple crystal structures for as many SuperCon entries as possible. Therefore, each SuperCon entry is paired with each crystal structure and the similarities between the chemical compositions are compared. In order to increase the number of matched entries, we normalize the chemical formulas before matching (see \ref{si:normalizing_chemical_formulas}) and perform artificial doping to approximate crystal structures of materials where the crystal structure of a very similar material is known.

\begin{figure}[th]
\centering
\includegraphics[width=\textwidth]{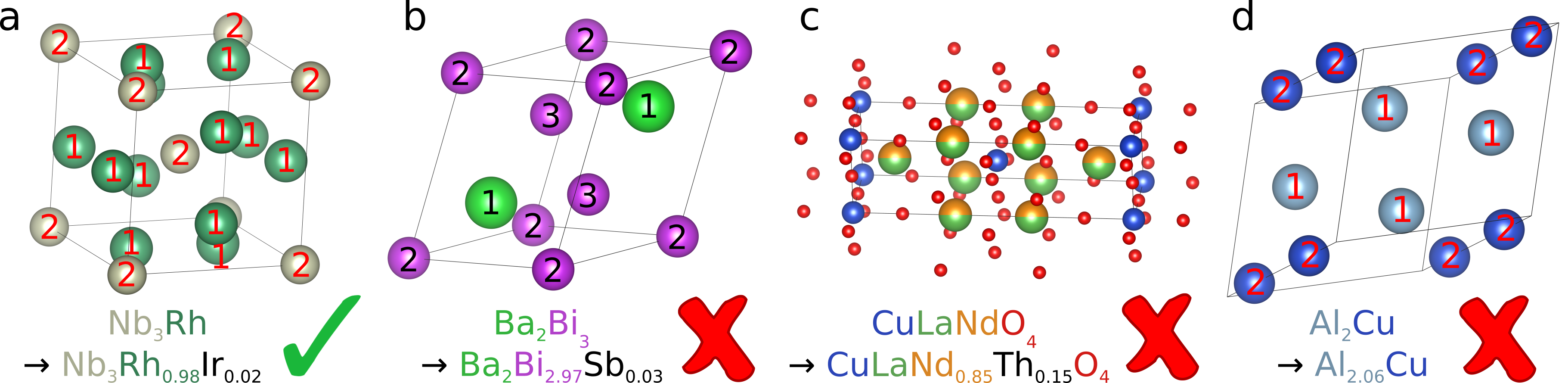}
\caption{Examples of SuperCon entries and respective candidate crystal structures before artificial doping. The checkmark shows if it is possible to use artificial doping to modify the chemical formula of the crystal structure (top) to fit the chemical formula of the SuperCon entry (bottom). The numbers on the atoms denote each set of symmetrically equivalent crystal sites. (a) shows a crystal with only one set of symmetrically equivalent Rh sites, (b) shows a crystal with two sets of symmetrically not equivalent Bi sites, (c) would generate a crystal structure with three elements on one site and (d) would require an additional crystal site.}
\label{fig:crystal_examples}
\end{figure}

\paragraph{Artificial doping:}
If the chemical formulas of SuperCon entry and crystal structure do not match perfectly but are still similar, we perform artificial doping. Artificial doping means that we use the crystal structure with a similar chemical formula as a proxy crystal structure for the real crystal structure of the SuperCon entry. We then partially replace the atoms at given crystal positions with other chemical elements, imitating real physical doping. After the replacement, the chemical formula of the new crystal structure matches perfectly the required chemical formula of the SuperCon entry. Note that this algorithm only changes the occupancies of the crystal sites. It does not change coordinates or interatomic distances. Besides, this algorithm can only be applied if the original chemical formulas are close enough, so that the real crystal structure of the SuperCon entry is likely to have similar crystal parameters (such as space group and lattice parameters) as the proxy crystal structure. Therefore, in order to keep the introduced bias small, we perform artificial doping only if the following requirements are met: 

\begin{enumerate}[label=(\alph*)]
\item \emph{The chemical formulas are similar:} We define three similarity metrics of chemical formulas, which are checked after normalizing the chemical formula of the crystal structure as explained above. These metrics are the absolute difference of atom numbers
\begin{equation}
\Delta_\mathrm{abs,i} = |x_\mathrm{sc,i} - x_\mathrm{cry,i}|, \label{eq:Delta_abs}	%(1)
\end{equation}
the relative difference of atom numbers
\begin{equation}
 \Delta_\mathrm{rel,i} = \frac{2 |x_\mathrm{sc,i} - x_\mathrm{cry,i}|}{x_\mathrm{sc,i} + x_\mathrm{cry,i}}, \label{eq:Delta_rel}%	(2)
\end{equation}
and the total weighted relative difference
\begin{equation}
 \Delta_\mathrm{totrel} = \frac{2 \Sigma_i |x_\mathrm{sc,i} - x_\mathrm{cry,i}|}{\Sigma_i x_\mathrm{sc,i} + x_\mathrm{cry,i}}. \label{eq:Delta_totrel}	%(3)
\end{equation}
$x_\mathrm{sc,i}$ and $x_\mathrm{cry,i}$ are the quantities of element $i$ of the chemical formulas of SuperCon entry and crystal structure, respectively. A pair of SuperCon entry and crystal structure is considered similar if $\Delta_\mathrm{abs,i} \leq 0.30 \quad \mathrm{OR} \quad \Delta_\mathrm{rel,i} \leq 0.20 \quad \forall i$ and $\Delta_\mathrm{totrel} \leq 0.15$, and the SuperCon entry has the same elements or up to one additional element as the crystal structure. One exception is that pure elements only match the same pure element. The thresholds were chosen to yield sensible results for a number of randomly selected examples. The exact threshold values only partially influence the final result, as only the most optimal matches according to additional criteria (see later) will be added to the final datasets. \autoref{tab:matching_examples} illustrates the procedure by showing examples of chemical formulas and whether they are considered similar, based on the definitions above.

The upper bound on $\Delta_\mathrm{rel,i}$ ensures that for each element, the relative difference of the chemical formulas is at most \SI{20}{\percent}. However, this requirement does not work well for doped materials such as \ce{CuLa_{1.95}Nd_{0.05}O4}. This chemical formula is close to one with a higher Nd doping concentration (e.g.\ \ce{CuLa_{1.90}Nd_{0.10}O4}), even though $\Delta_\mathrm{rel}$ is \SI{67}{\percent} for Nd. Therefore, the metric $\Delta_\mathrm{abs,i}$ allows for absolute differences of 0.3 or less for an element even though the requirement on $\Delta_\mathrm{rel}$ would be violated. The metric $\Delta_\mathrm{totrel}$ ensures that the \SI{20}{\percent} boundary is maxed out preferably for elements with a low number of atoms (and therefore low weight) in the chemical formula.

\item \emph{The necessary replacement of elements for artificial doping is unambiguous:} In a real crystal structure, not all crystal sites occupied by the same chemical element are equivalent. The dopant might prefer specific crystal sites due to differences in the local environment and thus free energy. Without further analysis, our artificial doping algorithm cannot determine which of the possible crystal sites becomes doped. Therefore, artificial doping is performed only if there is not more than one set of equivalent crystal sites for the dopants. In this case, the dopants are distributed equally over all equivalent crystal sites (see \autoref{fig:crystal_examples}). In \autoref{fig:crystal_examples}a, there is only one set of equivalent Rh sites. Therefore Rh can be unambiguously replaced with Ir. In \autoref{fig:crystal_examples}b there are two sets of equivalent Bi sites. It is not obvious which of these sites would be doped with Sb, therefore no artificial doping is performed.

We define equivalent crystal sites as all crystal sites which would have the same probability of being doped with a certain element, i.e.\ the sites are symmetrically equivalent or the sites are already doped or partially occupied by the same elements in the same quantities and therefore empirically behave identically under doping. The latter condition is important since a large number of cuprates in the ICSD would otherwise be excluded.

\item \emph{The replacement does not lead to crystal sites with more than two elements:} \autoref{fig:crystal_examples}c shows an example of this requirement. Even though the necessary replacement of elements would be unambiguous, we decided to discard such cases, so that each crystal site is doped with at most two different elements.

\item \emph{Artificial doping does not add or remove a crystal site:} Artificial doping is supposed to introduce only a minor bias. As such, it is acceptable to slightly modify the occupation numbers quantitatively, but fully removing a crystal site would constitute a severe change in the crystal structure. Furthermore, adding a crystal site is not possible since its position cannot be determined without further analysis. This is illustrated in \autoref{fig:crystal_examples}d.
\end{enumerate}

If requirements (a)-(d) are met, the appropriate quantity of the host element is replaced with the guest element at all equivalent crystal sites. In the ICSD, each element has its oxidation state given. When doping in a completely new element, its oxidation state is not known. In this case, we simply use the oxidation state of the host element for the new element.  Even with artificial doping not all SuperCon entries can be matched with a crystal structure. These entries are discarded.

\begin{table}
\setlength{\tabcolsep}{10pt}
\renewcommand{\arraystretch}{1.5}
\centering
\caption{Examples of pairs of chemical formulas of SuperCon entries and crystal structures. For each pair, the columns show the three similarity metrics (Equations \ref{eq:Delta_abs} - \ref{eq:Delta_totrel}).}
\label{tab:matching_examples}
\resizebox{\textwidth}{!}{%
\begin{tabular}{|l|l|c|c|c|l|}
\hline
\textbf{SuperCon} & \textbf{Crystal structure} & \multicolumn{1}{l|}{$\boldsymbol{\max(\Delta_\mathrm{abs})}$} & \multicolumn{1}{l|}{$\boldsymbol{\max(\Delta_\mathrm{rel})}$} & \multicolumn{1}{l|}{$\boldsymbol{\Delta_\mathrm{totrel}}$} & \textbf{Comment} \\ \hline
\ce{CuLa_{1.95}Nd_{0.05}O4} & \ce{CuLa_{1.90}Nd_{0.10}O4} & $0.05$ & $0.67$ & $0.01$ & \multicolumn{1}{p{0.5\textwidth}|}{\raggedright Matches after modification of the stoichiometry of La and Nd.} \\ \hline
\ce{CuLa_{1.95}Nd_{0.05}O4} & \ce{CuLa2O4} & $0.05$ & $2.0$ & $0.01$ & \multicolumn{1}{p{0.5\textwidth}|}{\raggedright Matching with modification. One site (La) of the unit cell is partially replaced by another element (Nd).} \\ \hline
\ce{CuLa2O4} & \ce{Cu2La4O8} & $0$ & $0$ & $0$ & \multicolumn{1}{p{0.5\textwidth}|}{\raggedright Matches perfectly (normalized chemical formulas are identical).} \\ \hline
\ce{NbGeC_{1.5}N_{0.5}} & \ce{NbGeC_{1.0}N_{1.0}} & $0.5$ & $0.67$ & $0.25$ & \multicolumn{1}{p{0.5\textwidth}|}{\raggedright Does not match because each metric is above its threshold (severe stoichiometry change would be required).} \\ \hline
\ce{W5Tc7} & \ce{W6Tc6} & $1$ & $0.18$ & $0.17$ & \multicolumn{1}{p{0.5\textwidth}|}{\raggedright Does not match because \totreldiff is above the threshold of $0.15$.} \\ \hline
\end{tabular}%
}
\end{table}

\paragraph{Keep only best matches:} The algorithm described above will generally match multiple crystal structures to each SuperCon entry. Therefore we identify the best matches by applying specific criteria: In the case of the Materials Project dataset, we first rank by the energy above hull \ehull (which is calculated in the Materials Project dataset for each crystal structure) and then by $\Delta_\mathrm{totrel}$. In both cases, lower values are preferred. If more than one crystal structure have the same optimal \ehull and $\Delta_\mathrm{totrel}$, both are kept in the database. In the case of the ICSD dataset, the ranking criteria is whether the crystal temperature is reported (preferred structures are the ones where the crystal temperature was given). If multiple crystal structures fulfill this criterion, all of them are added to our database. The ranking criteria were determined using hyperparameter optimization (see \ref{sec:sorting_criteria_optimization}).
The final 3DSC databases contain multiple crystal structures matched with the same SuperCon entry. This one-to-many mapping arises because multiple crystal structures might have the exact same rank after sorting. In the case of the \scmp, out of 5,759 SuperCon entries, only 14 are matched with each 2 crystal structures, so this is not a dominating issue. However in the case of the \scicsd, the in total 9,150 SuperCon entries are matched with 86,490 crystal structures (see statistical analysis in \autoref{sec:statistical_data_analysis}).

\section{Data Records}
\label{sec:data_records}

The \scmp dataset can be found under \url{https://github.com/aimat-lab/3DSC/tree/main/superconductors\_3D/data/final/MP/}. The directory 'cifs/' contains the cif files of all structures in the \scmp. The file '3DSC\_MP.csv' contains a table with all 5,759 entries in the \scmp.

The \scicsd can be found under \url{https://github.com/aimat-lab/3DSC/tree/main/superconductors\_3D/data/final/ICSD/}. Note that for the \scicsd, only the chemical formula and \Tc of each material as well as the ICSD ID of the original ICSD structure are given in the file '3DSC\_ICSD\_only\_IDs.csv' due to the restrictive ICSD license permissions. In order to generate the full \scicsd database, an ICSD license is required. Once the structures are downloaded, the matching and adaptation algorithm as described in our GitHub repository can be performed\cite{AimatlabSuperconductors3D}. 

The most important entries of the \scmp and the \scicsd are the chemical formula, the critical temperature in Kelvin and the path to the cif file which contains the corresponding three-dimensional crystal structure of each material. Non-superconductors are encoded as having a critical temperature of $\Tc = \kelvin{0}$. 
Additionally, the \scmp contains additional information from the original Materials Project dataset, e.g.\ electronic features derived from the original structures (before artificial doping) such as the band gap, the Fermi energy, the energy above hull or the total magnetization. The electronic and phonon density of states and band structures (if available in the Materials Project) are retrievable using task-IDs. 

The \scicsd in its full form (after re-running the matching and adaptation algorithm with structures from the ICSD) has similar entries as the \scmp. One difference is that structures from the ICSD do not contain any electronic features such as the band gap or the Fermi energy of the original structures. However, the \scicsd contains the crystal temperature \Tcry at which the structures were measured, which is missing in the \scmp. Since \Tcry was not reported for all of the structures and in doubt assumed to be room temperature (see \autoref{sec:data}), an additional binary entry indicates whether \Tcry was given explicitly in the ICSD or not.
Both datasets also contain entries which were important for the matching and adaptation algorithm and the analysis in this paper. These entries are important to simplify the reproduction and further work on improving the \scdataset. 
A more in-depth description of the entries and their exact names in the \scmp and the \scicsd can be found in the the \scdataset repository\cite{AimatlabSuperconductors3D}. 

\begin{figure}[t]
\centering
\includegraphics[width=0.78\textwidth]{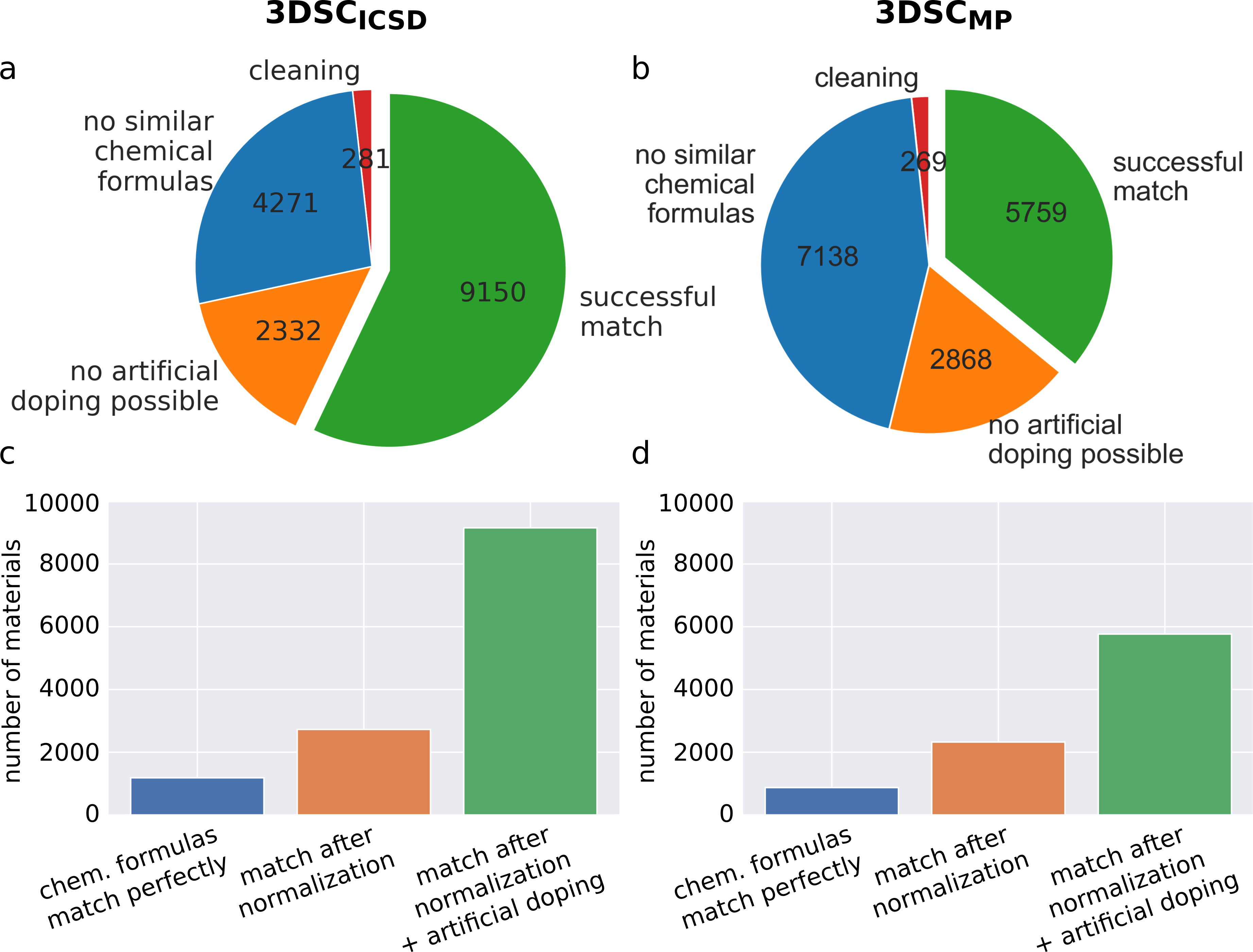}
\caption{Statistics of the matching algorithm. Panels a) and b) show the number of SuperCon entries that are lost in each step of the algorithm for the \scicsd and the \scmp dataset, respectively. Panels c) and d) show how many SuperCon entries in the final dataset were perfectly matched with crystal structures based on the absolute chemical formula, the normalized chemical formula, and how many were generated using the artificial doping algorithm.}
\label{fig:stats_matching_algo}
\end{figure}

\section{Results and discussion}

\subsection{Statistical data analysis}
\label{sec:statistical_data_analysis}

\begin{figure}[ht]
\centering
\includegraphics[width=0.7\textwidth]{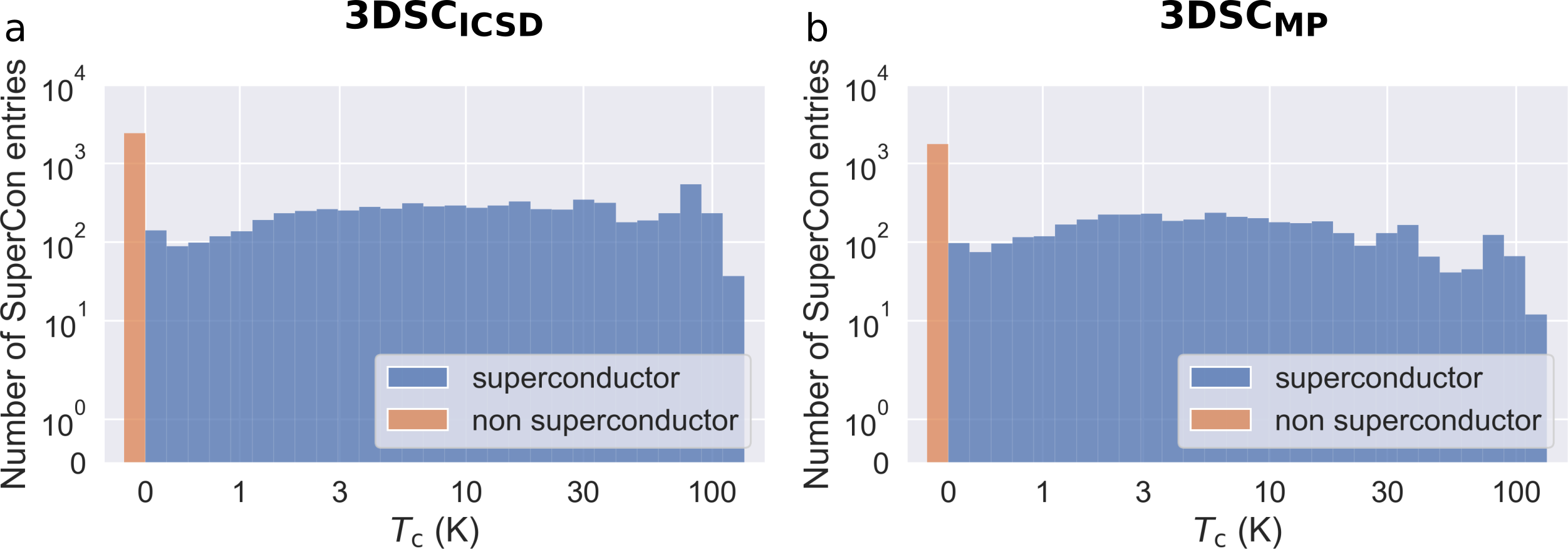}
\caption{The distribution of SuperCon entries per critical temperature \Tc for the \scicsd (a) and the \scmp (b).}
\label{fig:tc_hist_global}
\end{figure}

\autoref{fig:stats_matching_algo}a and b show the cleaning and matching statistics for the \scicsd and the \scmp, respectively. ‘No similar chemical formulas’ means that no crystal structure is close enough to be matched based on the metrics presented in \autoref{sec:matching_algorithm}. ‘No artificial doping possible’ means that artificial doping can not be performed for one of the other reasons explained in \autoref{sec:matching_algorithm}. As a result of the matching algorithm, approximately \SI{57}{\percent} of SuperCon entries can be matched with crystal structures from the ICSD and approximately \SI{36}{\percent} of SuperCon entries can be matched with structures from the Materials Project. Approximately \SI{93}{\percent} (5337) of the materials in the \scmp are also in the \scicsd. 

The bar plots in \autoref{fig:stats_matching_algo}c and d show how many material-crystal structure matches are obtained by performing the proposed matching algorithm in contrast to previously reported matching methods which only compare the absolute chemical formulas\cite{stanevMachineLearningModeling2018}. Matching normalized chemical formulas as well as performing artificial doping significantly increases the number of matched materials. While normalizing chemical formulas before matching is simple, it doubles the amount of SuperCon entries that can be matched for the \scicsd and even triples it for the \scmp. In addition, artificial doping roughly triples the matched materials for each dataset again. In contrast to standard matching, our proposed matching algorithm leads to a gain of \SI{773}{\percent} and \SI{660}{\percent} of matched SuperCon entries for the \scicsd and the \scmp respectively. 

When training machine learning models on datasets, an unbalanced distribution of labels can introduce bias, leading to systematic over- or underestimation of the critical temperatures in certain ranges. \autoref{fig:tc_hist_global}a and b shows the distribution of \Tc in the datasets for the \scicsd and the \scmp database, respectively. On a logarithmic scale, the number of superconductors per \Tc is relatively constant. One exception is the relatively high number of superconductors with a critical temperature of approximately $\Tc=\kelvin{90}$, which can be attributed to a widely studied class of superconductors based on \ybco which has a critical temperature of $\Tc = \kelvin{92}$. Elemental prevalence plots and further statistics of the distribution of \Tc, broken down into different groups of superconductors, can be found in \ref{si:statistics}.

Finally, it is important to evaluate the statistical influence of having multiple crystal structures per chemical formula in the \scicsd database. This is an issue mostly for the \scicsd since the \scmp dataset rarely has multiple structures per SuperCon entry. However, with a different choice of the sorting criteria, this issue would also exist for the \scmp since it is an inbuilt consequence of the matching approach. 

\begin{figure}[t]
\centering
\includegraphics[width=0.8\textwidth]{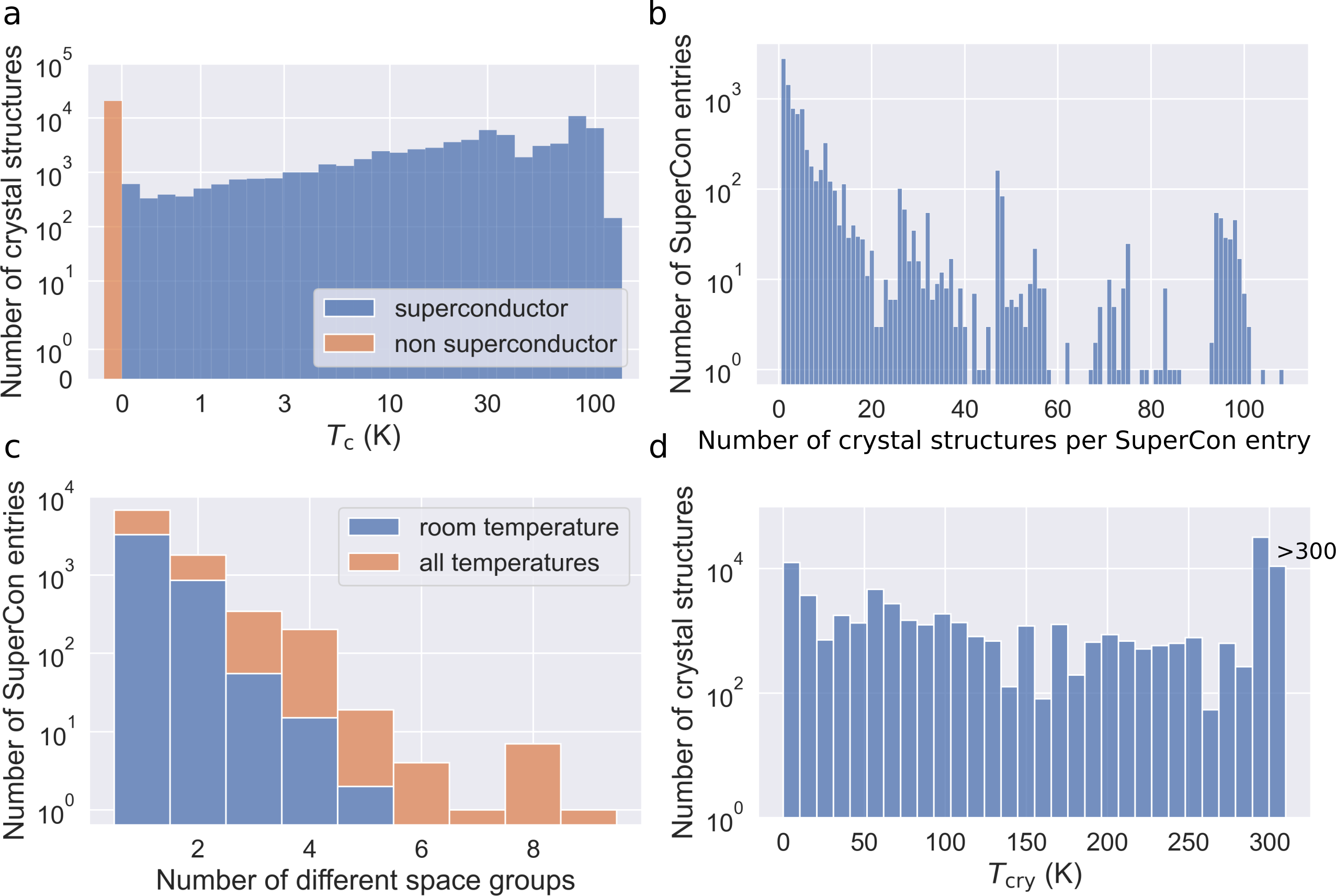}
\caption{Statistics regarding the mapping of one SuperCon entry to multiple crystal structures in the \scicsd. (a) shows a histogram of the number of crystal structures per \Tc. (b) shows the number of candidate crystal structures generated by the artificial doping algorithm per SuperCon entry. c) shows a histogram of the number of different space groups for each SuperCon entry, both for all crystal temperatures and only for structures at room temperature. (d) shows the number of crystal structures at a given crystal temperature \Tcry. For a better overview, all structures with a crystal temperature $\Tc > \kelvin{300}$ are collected in the last bar. \SI{50}{\percent} of the crystal structures have a crystal temperature $\Tcry < \kelvin{273}$.}
\label{fig:stats_icsd_add}
\end{figure}

\autoref{fig:stats_icsd_add}a shows \Tc dependent crystal structure counts in the dataset. For comparison, \autoref{fig:tc_hist_global}a shows the distribution of different superconductors instead of crystal structures. We find that the ratio of high-\Tc data points to low-\Tc data points has increased (ignoring the very last bin in the histogram). This shows that high-\Tc superconductors such as cuprates have matched with more crystal structures per material than low-\Tc superconductors. This might pose a potential issue because the effective weight of high-\Tc superconductors to low-\Tc superconductors has shifted from what it was before. To mitigate this issue, we have used a sample weight equal to the inverse of the number of crystal structures per material in all machine learning experiments.

Some materials have matched a lot of crystal structures as shown in \autoref{fig:stats_icsd_add}b. Note that there is an exponential decrease until approximately 20 crystal structures per SuperCon entry. Most data points beyond this number are artifacts due to series measurements of the same crystal with varying temperature. Furthermore, some materials have matched crystal structures with a large number of different space groups as shown in \autoref{fig:stats_icsd_add}c. The color coding shows that many of these different space groups are data points that are measured at room temperature. This potentially poses an issue, because a machine learning model trained on the data will “see” many different crystal structures with different space groups and the same chemical formula, which all have the same critical temperature \Tc. This problem is partially mitigated by having many crystal structures measured at low temperatures as shown in \autoref{fig:stats_icsd_add}d. Note that for a better overview, all of the data points with $\Tc > \kelvin{300}$ are collected in the last bar. We assume that the low-temperature crystal structures are helpful in predicting \Tc, because they are more likely to be the superconducting structures. However, the issue of having multiple very different crystal structures mapping to the same critical temperature in the \scicsd has to be kept in mind and will be discussed in \autoref{sec:limitations_and_perspective}. 

\subsection{Machine learning results}
\label{sec:machine_learning_results}

For all experiments in this section we have used a gradient boosting (XGB) model with standard hyperparameters of the xgboost scikit-learn API\cite{XGBoostDocumentation}. We used XGB because it turned out to be both more accurate and faster than (hyper-parameter optimized) densely connected neural network and random forest models. For the cross-validation we have used $n$ randomly repeated 80:20 splits ($n=25$ for \scicsd, $n=100$ for \scmp). In all our machine learning experiments, following Meredig et al.\cite{meredigCanMachineLearning2018}, we accounted for some extrapolation between train and test set to make the task more realistic: We grouped the materials in the train and test set by their chemical system, so that materials with the same chemical system are either all in the train set or all in the test set. The chemical system of a material is defined as the set of all chemical elements which make up the material (incl. dopants),  e.g.\ Ba-Cu-O-Y for \ce{YBa2Cu3O7}.

We computed the MSLE for each repetition and report the mean as well as the standard error of the mean of the MSLE values.The MSLE metric was also used as the loss function of each single model, due to the distribution of \Tc values in our dataset. Whenever possible, the same train-test splits were used in different experiments. Additionally, each crystal structure was given a sample weight of the inverse of the number of crystal structures for this SuperCon entry, so that the total weight for every SuperCon entry was the same. 

Before being fed into the XGB model, the critical temperature \Tc was approximately logarithmically scaled using $\Tc^\prime = \arcsinh(\Tc/\Tc^0)$ with $\Tc^0 = \kelvin{1}$.
To represent the chemical formula as a numerical vector we used MAGPIE features\cite{wardGeneralpurposeMachineLearning2016b}. To represent the crystal structures we developed disordered SOAP (DSOAP) features, an extension of SOAP features\cite{bartokRepresentingChemicalEnvironments2013} for disordered crystal structures, and some symmetry information, as explained in the supporting information \ref{si:DSOAP}. Additionally we concatenated the MAGPIE features of the chemical formula when representing the crystal structure.

\subsubsection{Importance of structural information}

\begin{figure}[ht]
\centering
\includegraphics[width=\textwidth]{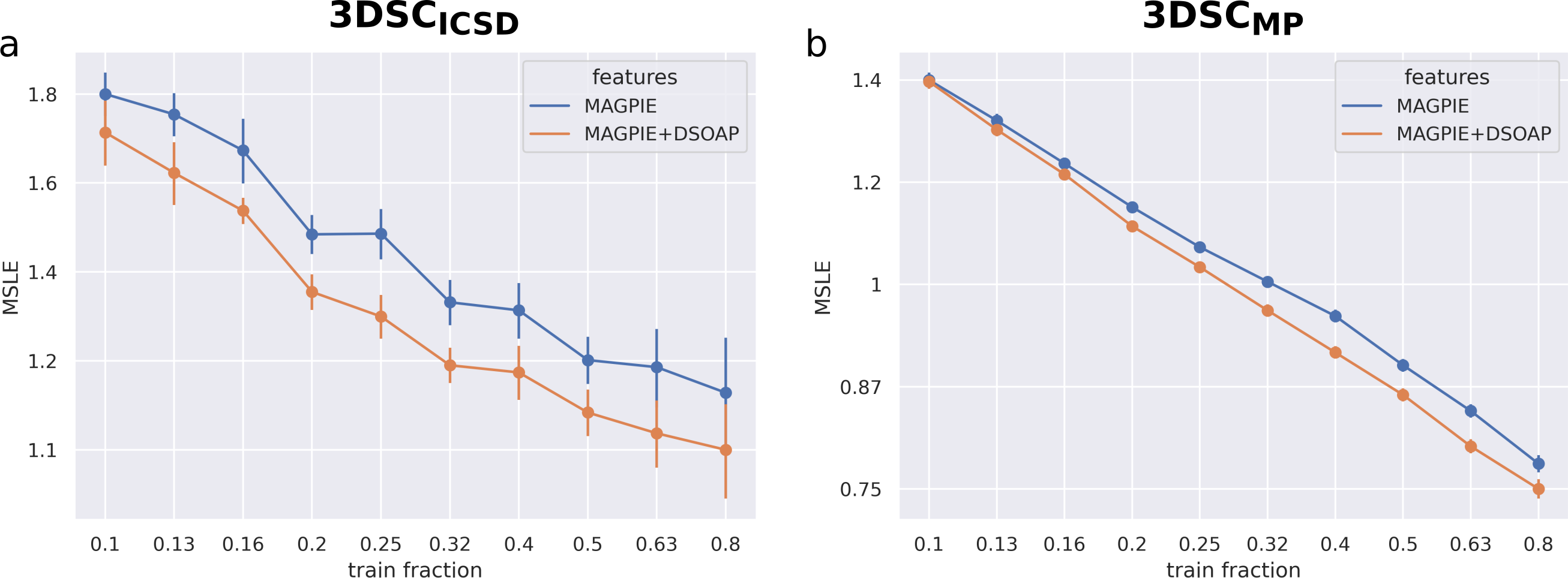}
\caption{A log-log plot of the learning curve of the \scicsd (a) and the \scmp (b) with and without access to structure-aware MAGPIE+DSOAP features. Note the different scales of the MSLE axes in (a) and (b).}
\label{fig:learning_curves}
\end{figure}

The performance of the XGB models trained with structural information on the \scicsd and the \scmp dataset is shown in \autoref{tab:sc_final_results}. Additionally, the performances on both datasets when trained only on the chemical formula is shown as reference. \autoref{fig:learning_curves} shows learning curves in which the performance of the XGB model with and without structural information is plotted for different train set sizes.

For both the \scicsd and the \scmp, training on the structural information improves the prediction of the critical temperature \Tc, despite the noise introduced by probably not always matching the correct structure. This is true even in the low-data regime as shown in \autoref{fig:learning_curves}. As an example, to convey a sense for the MSLE, a MSLE of $0.748$ for a superconductor with a true \Tc of \kelvin{1}, \kelvin{10} and \kelvin{100} corresponds to an absolute error of \kelvin{0.84}, \kelvin{4.36} and \kelvin{56.36}. In general the MSLE is lower for the \scmp than for the \scicsd. We assume that this is due to the fact that the \scicsd contains more doped materials. In such materials, slight changes in the feature space can correspond to large changes in \Tc. Therefore, the \scicsd might be harder to predict than the \scmp.

Furthermore, while the test error of the models with structural information is smaller, the train error is actually larger than when training on the chemical formula. This is a sign that including the structural information leads to less overfitting of the models. These results show that information about the 3D structure of the crystal structure is crucial for the prediction of the critical temperature, particularly for a better generalization.

A consequence of the matching algorithm is that the \scicsd and the \scmp contain less materials than the original SuperCon database from Stanev et al.\cite{stanevMachineLearningModeling2018}. As a reference we compare our new XGB model trained on the 3DSC database with DSOAP features (see Table \ref{tab:sc_final_results}) to XGB models trained on the full SuperCon data with MAGPIE features. To make the results comparable, we used exactly the same test sets as before. Additionally, to stay within the extrapolation setting, we removed materials with chemical systems which already occur in these test sets. The resulting MSLE on the test set is  $1.092 \pm 0.028$ for the \scicsd and $0.704 \pm 0.006$ for the \scmp. It is not a surprise that for the \scmp, these results are slightly better than when training on the \scmp (MSLE = 0.748), which contains only \SI{36}{\percent} of the materials in the full SuperCon. In contrast, the fact that models trained with structural information on the \scicsd perform better (MSLE = 1.085) than when trained on the full SuperCon with twice as much data shows how useful the structural information is. This shows the potential of data-driven approaches, given that enough data is published according to FAIR and AI-ready standards\cite{wilkinsonFAIRGuidingPrinciples2016, schefflerFAIRDataEnabling2022}.

\begin{table}
\setlength{\tabcolsep}{10pt}
\renewcommand{\arraystretch}{1.5}
\centering
\caption{The final results of the XGB models trained on the MAGPIE (“chem. formula”) and the MAGPIE+DSOAP features (“structure”) for the \scicsd and the \scmp. We report mean and standard error of the MSLEs of 25 and 100 randomly repeated 80:20 splits of the \scicsd and the \scmp respectively.}
\label{tab:sc_final_results}
\begin{tabular}{|l|l|c|c|}
\hline
\multicolumn{1}{|c|}{\textbf{Dataset}} & \multicolumn{1}{c|}{\textbf{Features}} & \textbf{MSLE (test)} & \textbf{MSLE (train)} \\ \hline
\multirow{2}{*}{\scicsd} & Chem. formula & $1.176 \pm 0.095$ & $0.155 \pm 0.005$ \\ \cline{2-4} 
 & Structure & $\mathbf{1.085 \pm 0.073}$ & $0.275 \pm 0.008$ \\ \hline
\multirow{2}{*}{\scmp} & Chem. formula & $0.776 \pm 0.010$ & $0.078 \pm 0.001$ \\ \cline{2-4} 
 & Structure & $\mathbf{0.748 \pm 0.010}$ & $0.135 \pm 0.001$ \\ \hline
\end{tabular}
\end{table}

\subsubsection{Importance of  structural information for different groups of superconductors}
\label{sec:physical_groups}

\begin{table}
\setlength{\tabcolsep}{10pt}
\renewcommand{\arraystretch}{1.5}
\centering
\caption{The number of different materials in the \scicsd and the \scmp for each group of superconductors.}
\label{tab:num_sc_groups}
\begin{tabular}{|l|c|c|}
\hline
\multicolumn{1}{|c|}{\textbf{Group}} & $\mathrm{\mathbf{3DSC_{ICSD}}}$ & $\mathrm{\mathbf{3DSC_{MP}}}$ \\ \hline
Others & 4068 & 3525 \\ \hline
Cuprates & 3189 & 874 \\ \hline
Ferrites & 936 & 517 \\ \hline
Heavy fermions & 402 & 418 \\ \hline
Oxides & 384 & 310 \\ \hline
Chevrel phases & 103 & 74 \\ \hline
Carbon-based & 46 & 30 \\ \hline
\end{tabular}
\end{table}

\begin{figure}[ht]
\centering
\includegraphics[width=\textwidth]{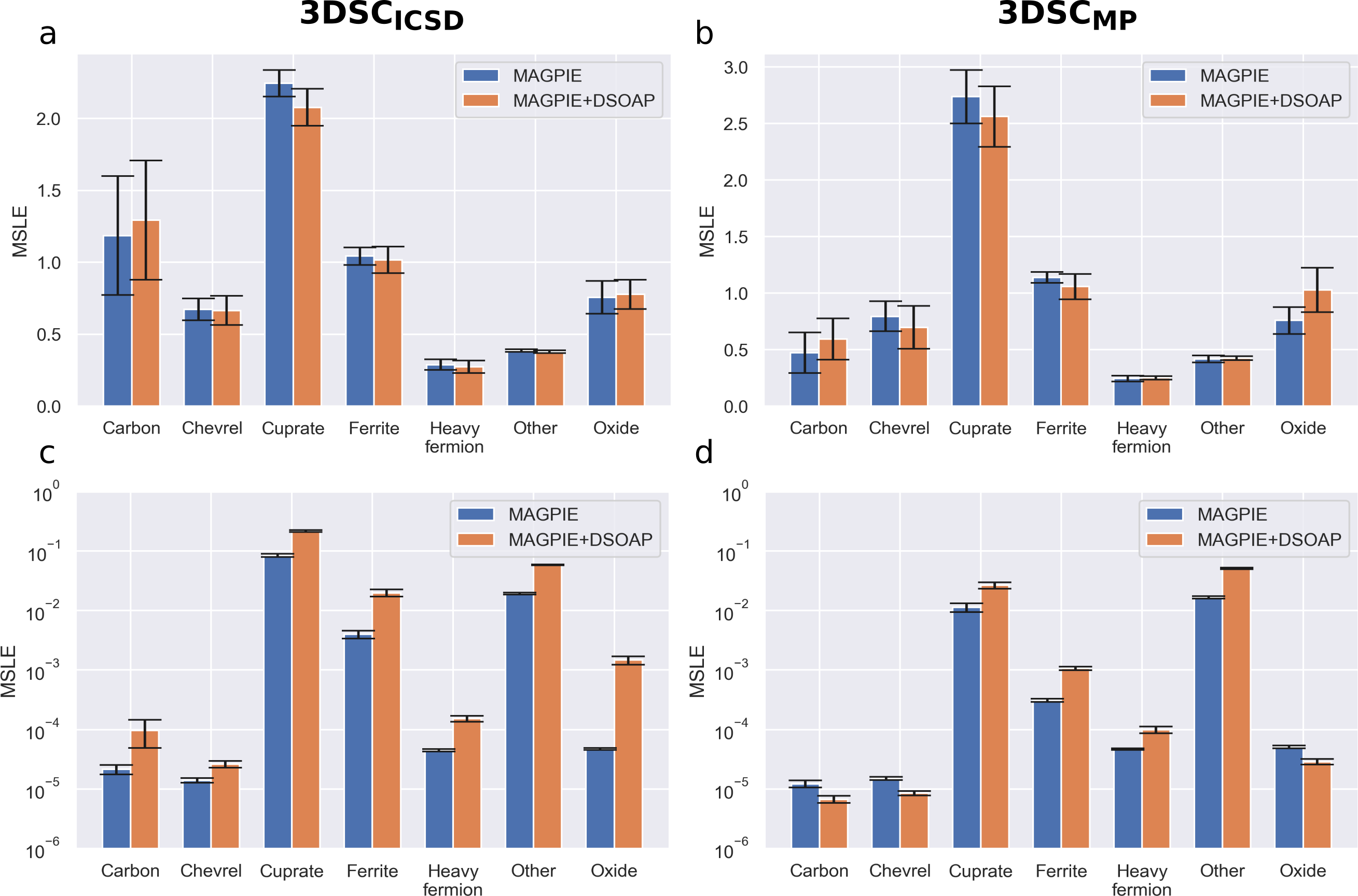}
\caption{The results of training and testing separately for each superconductor group. For comparison we show the results with XGB models trained only on MAGPIE features and trained on MAGPIE+DSOAP features. Shown are mean and standard error of the MSLEs obtained in a 5-fold cross validation grouped by chemical system. The first row shows the performance on the test set for the \scicsd (a) and the \scmp (b). The second row shows the performance on the train set for the \scicsd (c) and the \scmp (d).}
\label{fig:only_physical_groups}
\end{figure}

The SuperCon dataset is very clustered, with many data points coming from narrow groups of materials. Having access to the 3D crystal structure information might influence different superconductor groups in different ways. In order to test the performance within different groups of superconductors with and without structural information, we have partitioned the \scicsd and the \scmp dataset into seven different datasets containing only one class of superconductors each, namely cuprates, ferrites, heavy fermion materials, Chevrel phases, oxides and carbon-based materials. All following models are trained and tested on only one group of superconductors each. This grouping is based on the chemical formula and is done automatically when cleaning the SuperCon dataset in the matching and adaptation algorithm. If a data point was attributed to no group, it was added to a group “others”, and if it was attributed to multiple groups, it was excluded in this analysis. The number of materials in each group and for each dataset is shown in \autoref{tab:num_sc_groups}.

To compare the influence of the structural features for each group independently, we trained XGB models with 5-fold cross validation grouped by chemical system (see \autoref{sec:machine_learning_results}) on each group, once only with MAGPIE features, encoding only the chemical formula, and once with MAGPIE+DSOAP features, encoding the crystal structure. The results on the test and train set are shown in \autoref{fig:only_physical_groups}. For the test error, the influence of the structurally aware MAGPIE+DSOAP features is not the same across all groups. Overall, the difference in performance between MAGPIE and MAGPIE+DSOAP features is always smaller than the standard error, which makes the difference not statistically significant. However, one can still analyze some trends: The cuprates are the group where the experiments with structural information have the most advantage compared to the runs without structural information. For ferrites, the runs with structural information are a bit better as well. Furthermore, the train error with structural features is higher than the train error without structural features for all groups of the \scicsd and for the bigger groups in the \scmp. This indicates that the structurally aware features are better for generalizing to unseen materials.

What is more interesting is the fact that some of the experiments with structural information have a worse test error than the runs without structural information, in particular for the carbon based materials and the oxides. We suspect that one reason for this behavior is the way the \scdataset is created, i.e.\ by matching chemical formulas of materials with the corresponding chemical formulas of crystal structures. One can imagine that such an approach fails for groups such as carbon-based materials, where the normalized chemical formula can look quite similar for very different 3D structures. Technically, it would be most useful for exactly these groups to have the correct structural information. Yet, the way the matching algorithm works, it is also likely to simply match very wrong structures. Another reason might be that at a given dataset size, adding additional features (8000 DSOAP features compared to 145 MAGPIE features) might actually increase overfitting and potentially decrease test set performance. The benefits of additional features will only become statistically significant once a certain training set size is reached (due to steeper learning curves)\cite{vonlilienfeldRetrospectiveDecadeMachine2020}. Such an overfitting effect can be observed for the smaller groups of the \scmp (carbon-based materials, Chevrel phases and oxides) where the train error decreases when adding structurally aware features while the test error is increases. 

For the cuprates, the situation is different: The chemical formula of the structure determines the structure quite well, so the matched structures are likely to be close to the real structures. This might be the reason why the prediction of the \Tc of cuprates works out better with MAGPIE+DSOAP features than only with MAGPIE features. 

Overall different groups of superconductors are influenced differently by adding the structural features, even though no clear conclusions can be drawn due to the small dataset sizes. One potential limitation that one should keep in mind is the fact that the matching algorithm might fail just for the groups which would most strongly benefit from structural information, emphasizing again the importance of reporting crystal structures and additional data in a machine readable and FAIR way. 

\subsection{Sensitivity analysis}

We have performed experiments to investigate the influence of two important parameters in the matching algorithm, the \maxtotreldiff and the normalization of the chemical formulas before matching. The plots and a more detailed discussion of these experiments can be found in the supporting information in \ref{si:sensitivity_analysis} for the \maxtotreldiff and in \ref{sec:importance_of_normalization} for the normalization of the chemical formulas. The final \maxtotreldiff used above was chosen to maximize the number of matched materials while minimizing the bias introduced by artificial doping. Our results furthermore show that normalizing the chemical formula before matching and artificial doping is beneficial, indicating that many entries in the SuperCon database do not reflect the exact unit cell. Additionally, we implemented a stricter version of the artificial doping algorithm, where only one single \scicsd crystal structure is selected for each superconductor, rather than all matching structures. We found that the mean performance becomes slightly worse, but the shift was not statistically significant.

\section{Limitations and perspective}
\label{sec:limitations_and_perspective}

The main limitation of the matching algorithm and the \scdataset is that there is no guarantee that a matched crystal structure is the correct superconducting structure for this material. This is particularly obvious for the \scicsd where there can be up to 9 different space groups for the same SuperCon entry. The \scicsd tries to mitigate this problem by ‘diluting’ uninteresting structures with more interesting structures. The \scmp tries to counter this problem by using the more stable structures identified by the energy above hull. Ultimately, this problem can only be solved by manually choosing the correct superconducting structure based on expert knowledge or measuring the crystal structure at temperatures close to \Tc in order to find the correct crystal phase.

Until the ranking procedure, the matching and adaptation algorithm is very general and tries to keep as much information as possible. Only the step of selecting which structures are most likely to be the superconducting ones is based on empirical assumptions and thus introduces bias and potentially noise. We therefore expect further improvements by adjusting the sorting criteria for ranking crystal structures. Two possible additional sorting criteria are the crystal temperature reported in the ICSD database, and space groups reported for some of the SuperCon entries. The idea behind sorting according to crystal temperature is that structures measured at lower temperatures are more likely to be the superconducting structure. However, this approach might lead to an artificially introduced correlation between \Tc and \Tcry. Therefore we decided to not include this criteria in our work. The most likely space group for each material can be identified either by checking the sparse structural information in the original SuperCon database or by checking ICSD structures for keywords regarding superconductivity in the paper titles or abstracts. However, this approach only covers a small fraction of materials.

So far, the \scdataset focuses on adding structural information. A promising addition would be to add electronic information, e.g.\ from the Materials Project into the \scmp. Examples include the electronic structure, the band gap, the total energy, the formation energy and the Fermi energy, which are given for more than \SI{90}{\percent} of the materials in the Materials Project. However, the equivalent of artificial doping for these electronic features would at least require additional quantum chemical calculations, if possible at all (e.g.\ for materials with small doping concentrations).

In this work, we have refrained from merging the \scmp and the \scicsd, since already \SI{93}{\percent} of the materials in the \scmp are in the \scicsd and the resulting database could have not been published freely. For future work, it might be interesting to expand the \scmp using other publicly available datasets such as the Crystallography Open Database\cite{grazulisCrystallographyOpenDatabase2009} (COD) or the Automatic FLOW for Materials Discovery\cite{curtaroloAFLOWLIBORGDistributed2012} (AFLOW) database to maximize the number of materials in the \scdataset. 

\section{Summary and conclusion}

We have created two datasets, which contain the critical temperature \Tc and approximated 3D crystal structures of 9,150 (\scicsd) and 5,759 (\scmp) superconducting and non-superconducting materials. The datasets are built on a public version of the SuperCon database, enriched with crystal structures which were generated from the Materials Project and the ICSD database, using an artificial doping algorithm presented in this paper. We publicly provide the full \scmp dataset, as well as the ICSD IDs of each material in the \scicsd. Additionally we provide the code we developed to create the databases, to enable recreation and extension of the datasets. We demonstrated that the additional information provided by the crystal structures leads to a performance gain of machine learning models trained on our datasets. We furthermore discussed the limitations of the current \scdataset and how they could be approached in future work. 

Overall, our work demonstrates the added value of having access to FAIR and machine readable data on superconducting materials in publicly available databases. We hope that this motivates the scientific community working on superconductivity to publish their research data in public databases, including structural information as well as additional meta-data.

\section*{Usage Notes}

We provide the full code used for the dataset generation and the analysis in this paper. In order to simplify reproduction and further work on the \scmp, we provide a single python script to generate the \scmp, plot most of the statistical plots and generate the learning curves. 
Note that due to memory constraints, the raw \scmp file on GitHub is missing the DSOAP and MAGPIE features which we used for the machine learning experiments in this paper (see \autoref{sec:machine_learning_results}). Re-running the matching and adaptation algorithm will also include these feature vectors.
Future efforts in training machine learning models on the \scmp can be based on the provided datasets, in particular the readily-available \scmp. Future efforts on improving the \scdataset can be based on the provided code.  

\section*{Code availability}

Code and data are available free of charge on \url{https://github.com/aimat-lab/3DSC}.

\section*{Acknowledgements}

This work was performed on the computational resource bwUniCluster funded by the Ministry of Science, Research and the Arts Baden-Württemberg and the Universities of the State of Baden-Württemberg, Germany, within the framework program bwHPC.

\bibliography{sample}

\section*{Author contributions statement}

T.S. conceived the database, wrote the code and conducted the experiments, T.S. and P.F. conceived and analyzed the experiments, R.W. and J.S. contributed expert knowledge regarding superconductors and contributed to the planning and analysis of the experiments, P.F. supervised the project. All authors contributed to the manuscript.

\section*{Competing interests}

The authors declare no competing interest.

\renewcommand{\appendixname}{Supplementary Information}

\appendix
\renewcommand{\thesection}{S\arabic{section}}

\newpage
%\section*{Supplementary Information}
\renewcommand{\thefigure}{S\arabic{figure}}
\setcounter{figure}{0}
\renewcommand{\thetable}{S\arabic{table}}
\setcounter{table}{0}

\section{Normalization of chemical formulas}
\label{si:normalizing_chemical_formulas}

Instead of matching chemical formulas only when they match exactly, we also match chemical formulas if they differ only by a constant factor. I.e. the SuperCon entry \ce{CuLa2O4} would also be matched by a crystal structure with the chemical formula \ce{Cu2La4O8} with a relative factor of $1/2$. This increased the matched entries by a large factor, which is why we assume that sometimes the experimental authors of the SuperCon data did not know the exact composition of the unit cell and just wrote down the smallest ratio of integers. We implemented this by multiplying the chemical formula of the crystal structure with the ratio of the sum of the atom numbers of the two chemical formulas.

\section{Additional dataset statistics}
\label{si:statistics}

\begin{figure}[pht]
\centering
\includegraphics[width=\textwidth]{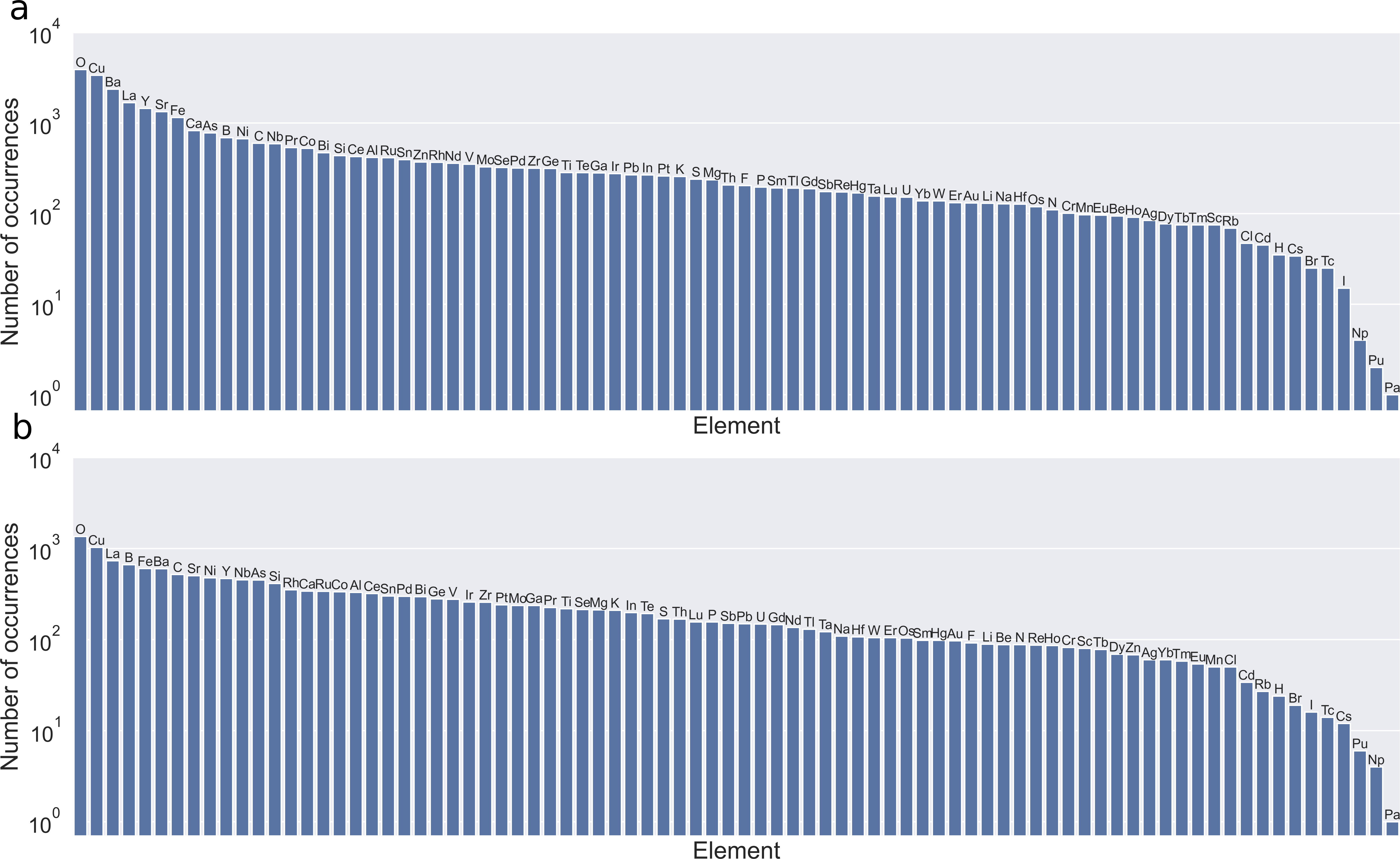}
\caption{These elemental prevalence plots show how often each chemical element occurs in the \scicsd (a) and the \scmp (b).}
\label{fig:stats_dataset}
\end{figure}

\begin{figure}[pht]
\centering
\includegraphics[width=\textwidth]{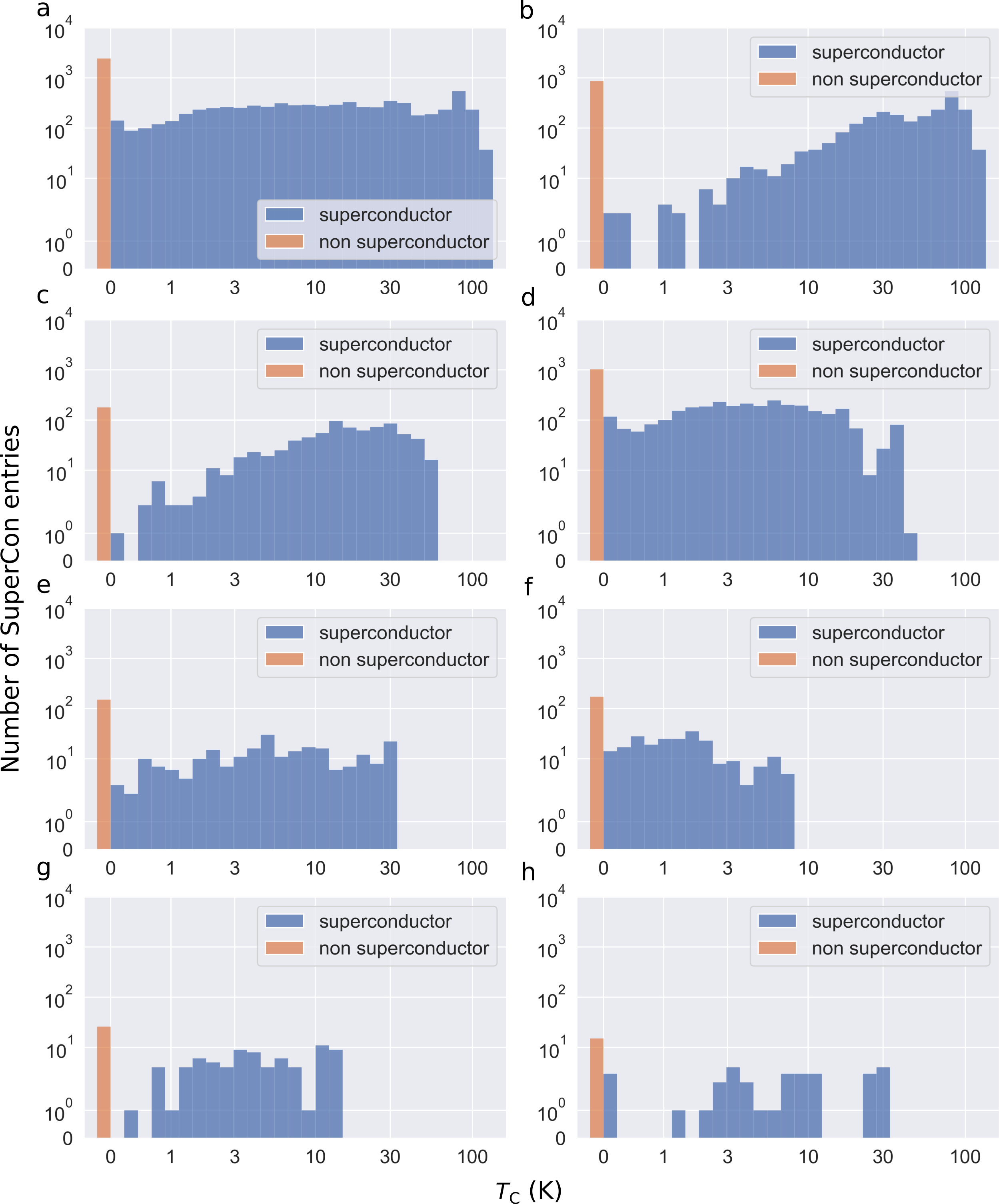}
\caption{The number of SuperCon entries with given critical temperature \Tc for the full \scicsd (a) and for each superconductor group ((b) to (h)): cuprates (b), ferrites (c), other (d), oxide (e), heavy fermion materials (f), Chevrel phases (g), carbon based materials (h). The non-su\-per\-con\-duc\-tors are shown in orange. The left-most blue bar includes only superconductors with $\Tc > \kelvin{0}$.}
\label{fig:stats_icsd_tc_hist}
\end{figure}

\begin{figure}[pht]
\centering
\includegraphics[width=\textwidth]{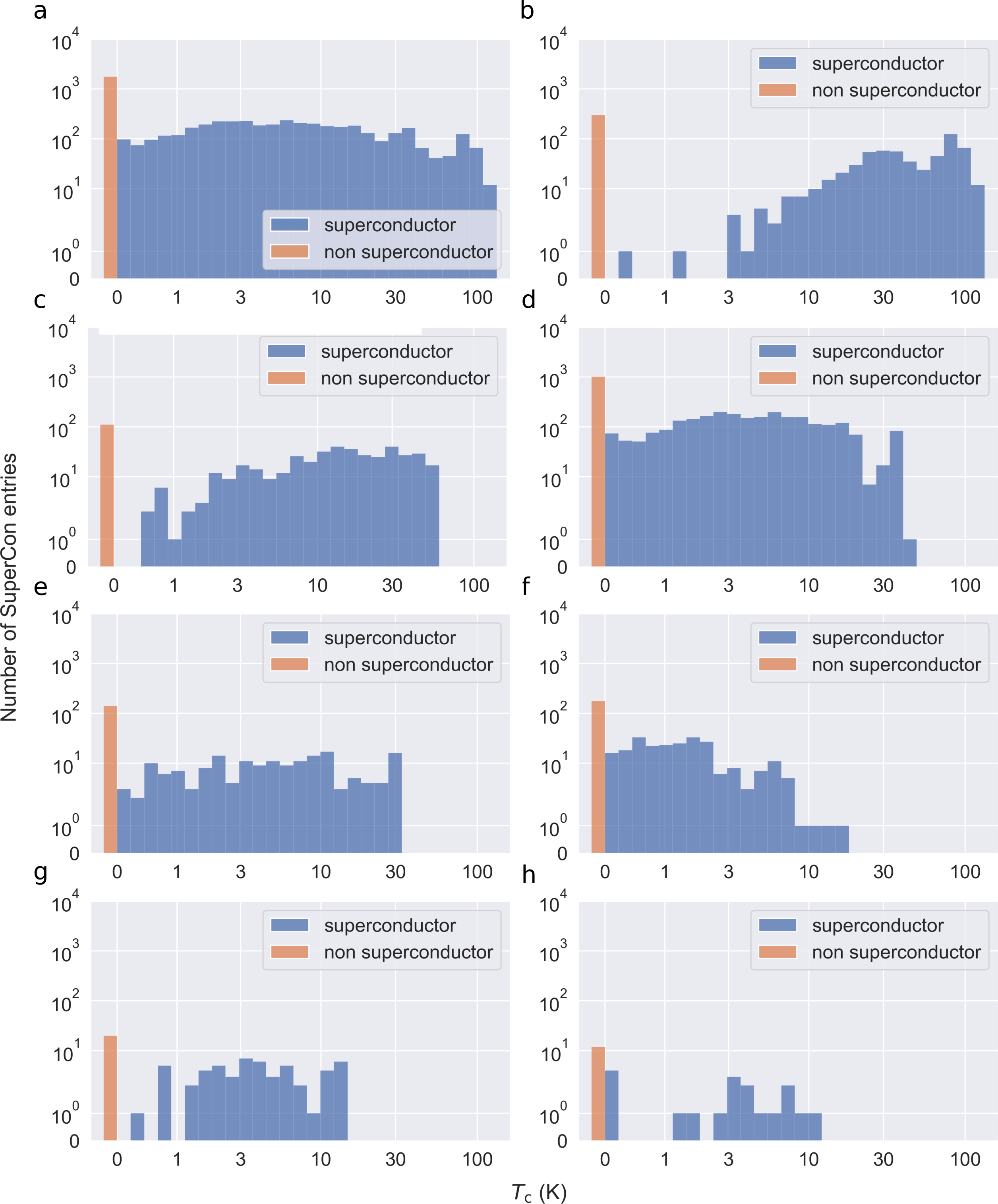}
\caption{The number of SuperCon entries with given critical temperature \Tc for the full \scmp (a) and for each superconductor group ((b) to (h)): cuprates (b), ferrites (c), other (d), oxide (e), heavy fermion materials (f), Chevrel phases (g), carbon based materials (h). The non-su\-per\-con\-duc\-tors are shown in orange. The left-most blue bar includes only superconductors with $\Tc > \kelvin{0}$.}
\label{fig:stats_mp_tc_hist}
\end{figure}
\newpage

\section{Disordered SOAP (DSOAP) features}
\label{si:DSOAP}

\begin{figure}
\centering
\includegraphics[width=0.25\textwidth]{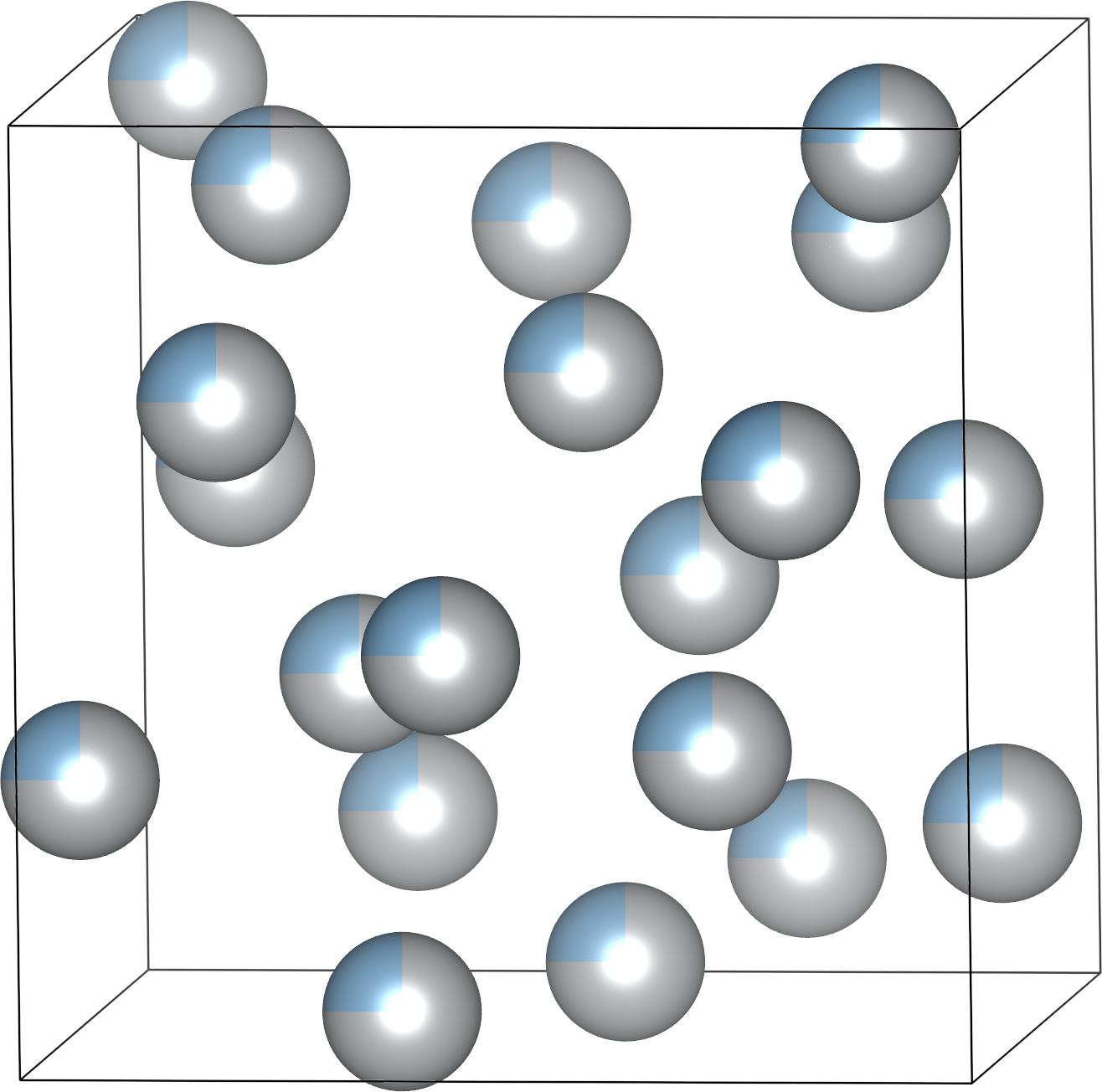}
\caption{This structure has 20 doped crystal sites in its primitive unit cell, but they all belong to only 2 different sets of symmetrically equivalent crystal sites.}
\label{fig:Ag0.75Al0.25}
\end{figure}

To represent the 3D crystal structures for machine learning algorithms, we chose SOAP features\cite{bartokRepresentingChemicalEnvironments2013} which are calculated using the python package Dscribe\cite{himanenDScribeLibraryDescriptors2020}. Yet, these features only support ordered crystal structures, but not disordered structures with fractional occupancies such as vacancies and doping. Still, such structures frequently appear in the \scdataset. To address this issue we have generalized the SOAP features to ‘Disordered SOAP’ (DSOAP) features. These DSOAP features are based on the original SOAP features of ordered structures, but they also incorporate information such as doping and vacancies. 

We will now explain the implementation of the DSOAP features and give an example. The idea behind DSOAP features is simple: Each disordered structure is understood as consisting of a superposition of ordered structures. The weights in this superposition are given by the occupancies. The SOAP features of the ordered structures can be calculated with the Dscribe library. The DSOAP vector of the disordered structure is then given as a weighted average of the SOAP vectors of the ordered structures, with weights given by the occupancies.

\begin{sloppypar}
As an example we will discuss the DSOAP features for an (exemplary) crystal with chemical formula \ce{Fe_{0.35}Zn_{0.35}Te_{1.6}Se_{0.4}O_{0.5}}. Assume this crystal structure has four atom sites in its primitive unit cell:\\
\end{sloppypar}
\begin{tabular}{ll}
1. \ce{O_{0.5}} 			& (vacancy) \\
2. \ce{Te}				    & (ordered) \\
3. \ce{Te_{0.6}Se_{0.4}}	& (doped) \\
4. \ce{Zn_{0.35}Fe_{0.35}}	& (doped vacancy)\\
\end{tabular}

In the following we will explain each step to calculate the DSOAP features of this crystal structure:
\begin{enumerate}
\item \textbf{Vacancies:} For vacancies, the total occupancy of each atom site is recorded to use it as weight for this atom site. In this example, the recorded vacancy weights for the four sites are (0.5, 1, 1, 0.7). After recording these weights the occupancies of the vacancies are scaled so that afterwards each crystal site has a total occupancy of $1.0$. The four atom sites now have the following occupancies:\\
\begin{tabular}{ll}
1. \ce{O} 			        & (scaled by a factor of $1/0.5$) \\
2. \ce{Te}			        & (unchanged) \\
3. \ce{Te_{0.6}Se_{0.4}}	& (unchanged) \\
4. \ce{Zn_{0.5}Fe_{0.5}} 	& (scaled by a factor of $1/0.7$)     
\end{tabular}

\item \textbf{Doping:} For doped crystal structures, we generate all possible combinations of ordered structures that can arise from the doped elements. These ordered structures serve as proxy structures. Since there are two doped crystal sites, namely \ce{Te_{0.6}Se_{0.4}} and \ce{Zn_{0.5}Fe_{0.5}}, with two elements each, there are four proxy structures: \ce{FeTe2O}, \ce{FeTeSeO}, \ce{ZnTe2O}, \ce{ZnTeSeO}.

Additionally the doping occupancies are recorded as weights for each proxy structure. The weights for all crystal sites of a structure are multiplied to simulate the probabilities. The doping weights for each of the four proxy structures are:\\
\begin{tabular}{ll}
\ce{FeTe2O}: 	& $0.5 \cdot 1 \cdot 0.6 \cdot 1 = 0.3$ \\
\ce{FeTeSeO}: 	& $0.5 \cdot 1 \cdot 0.4 \cdot 1 = 0.2$ \\
\ce{ZnTe2O}:	& $0.5 \cdot 1 \cdot 0.6 \cdot 1 = 0.3$ \\
\ce{ZnTeSeO}: 	& $0.5 \cdot 1 \cdot 0.4 \cdot 1 = 0.2$ 
\end{tabular}

Note that they add up to 1, because we calculated the doping weights for the manipulated structure without vacancies.\\
\emph{Remark:} Because of the combinatorial nature of this algorithm, it can happen that the number of proxy structures becomes extremely high and computationally not tractable when doing this for each crystal site. To alleviate this problem, we calculate the combinations of proxy structures not per crystal site but per symmetrically equivalent set of crystal sites. An example for this issue (taken from the ICSD) is shown in \autoref{fig:Ag0.75Al0.25}: The crystal structure of \ce{Ag_{0.75}Al_{0.25}} has 20 doped crystal sites, but they all belong to only two different sets of symmetrically equivalent crystal sites. Therefore, instead of having to compute $2^{20}$ different ordered proxy structures we only need to compute $2^2$.

\item \textbf{SOAP features:} We calculate the SOAP vectors of each crystal site of all proxy structures with the Dscribe library. This is no issue anymore since these structures are completely ordered. Since in our example there are 4 proxy structures with 4 crystal sites each, there are 16 SOAP vectors in total.

\item \textbf{Weighted average of crystal sites:} We compute the full SOAP vector of one proxy structure by doing a weighted average over all of its crystal sites. The weights are the recorded vacancy weights, i.e.\ the original total occupancy of each crystal site. In this example these weights are (0.5, 1, 1, 0.7). The weighted average $\boldsymbol{v}$ of some vectors $\bm{v}_i$ with weights $w_i$ is defined as 
\begin{equation}
    \boldsymbol{v}(w_i,\bm{v}_i) = \frac{\sum_i w_i \bm{v}_i}{\sum_i w_i} \label{eq:weighted_average}
\end{equation}

\item \textbf{Weighted averages of ordered structures:} We compute the DSOAP vector of the original disordered structure by doing a weighted average over all of the ordered proxy structures. The weights are the recorded doping weights. In this example these weights are (0.3, 0.2, 0.3, 0.2).
\end{enumerate}
Note that the order of the weighted averages doesn't matter since it is just two times a linear combination of vectors after each other:
\begin{equation}
    \bm{c} = \boldsymbol{v}_\sub{dop}(\boldsymbol{v}_\sub{vac}(\bm{s}_{ij}^\sub{proxy})) = \sum_{i,j} \frac{w_i ^\sub{dop} w_j ^\sub{vac}}{(\sum_k w_k ^\sub{dop}) \cdot (\sum_l w_l ^\sub{vac})} \bm{s}_{ij} ^\sub{proxy}
\end{equation}
where $\bm{c}$ is the DSOAP vector for the final crystal, $w_i ^\sub{dop}$ and $w_j ^\sub{vac}$ are the weights for the doping and the vacancies and $\bm{s}_{ij} ^\sub{proxy}$ is the SOAP vector of the $j$th crystal site of the $i$th proxy structure.

Note also that if one inputs an ordered crystal structure, the output is reduced to the original SOAP features.

Additionally to the DSOAP features we used symmetry features \fsym. This was done simply because we have this symmetry given automatically and we assumed it to be helpful for predicting \Tc if the symmetries would be encoded explicitly. These symmetry features had 11 entries: The first 7 entries encoded the 7 crystal systems (cubic, hexagonal, monoclinic, orthorhombic, tetragonal, triclinic, trigonal) with the corresponding point group encoded as an integer in these 7 feature vectors. Additionally the bravais-centring (primitive, base-centered, body-centered, face-centered) is one-hot encoded as 4 additional binary features. In all experiments these \fsym features are always implicitly appended when using DSOAP features. However, our analysis of the results showed that the symmetry features made our results slightly better in the case of the \scicsd and slightly worse in the case of the \scmp, suggesting that the information added by \fsym features is limited (see \ref{si:random_dropping_and_fsym}).

\section{Additional machine learning experiments}
\label{si:sensitivity_analysis}

\subsection{The difference of chemical formulas \totreldiff}

\begin{figure}
\centering
\includegraphics[width=\textwidth]{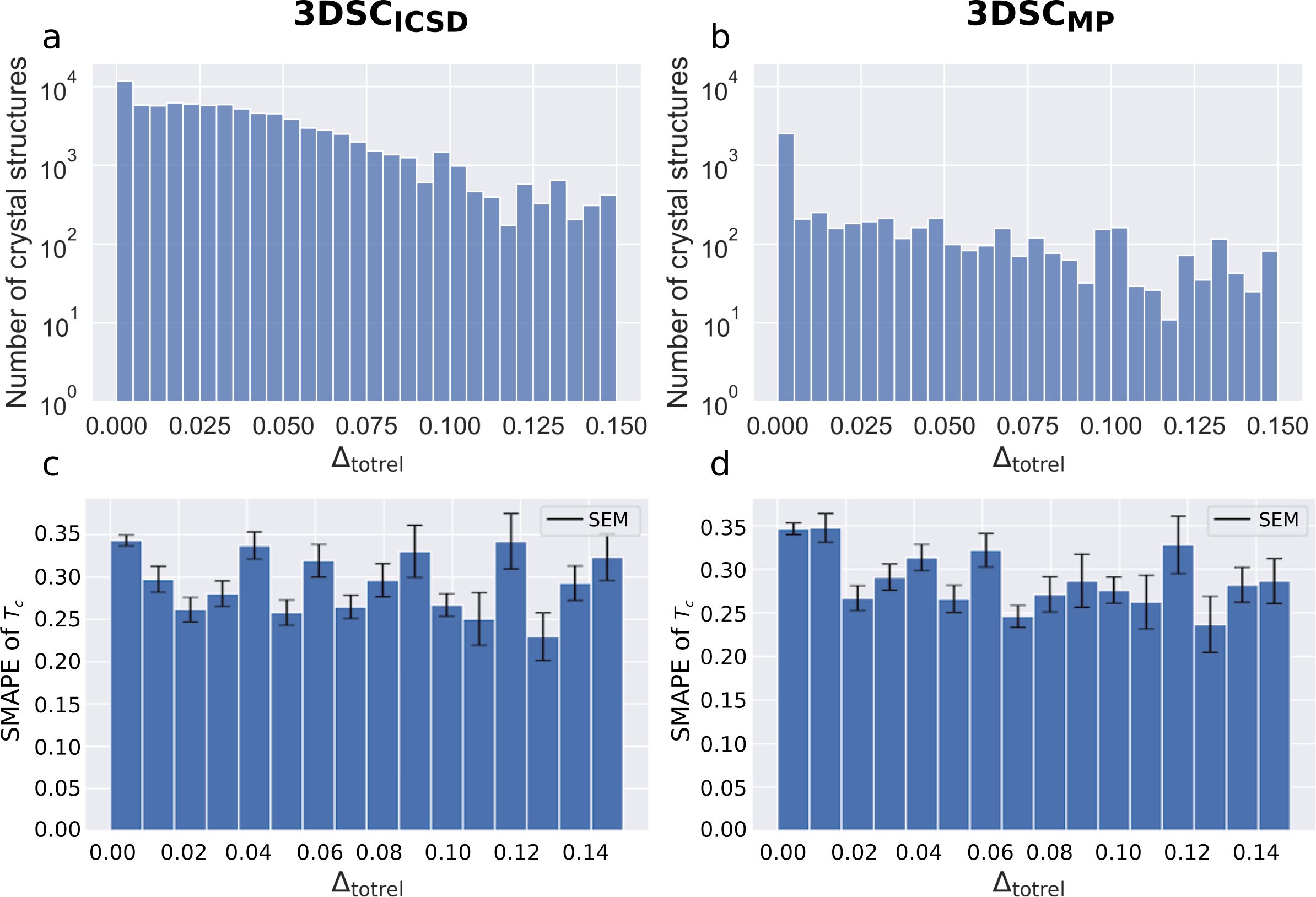}
\caption{Plots regarding the role of the parameter \totreldiff. The first row shows the number of crystal structures with a given \totreldiff between the chemical formula of the SuperCon entry and the original crystal structure for the \scicsd (a) and the \scmp (b). The second row shows the Symmetric Mean Absolute Percentage Error (SMAPE) and the standard error of the mean (SEM) of the critical temperature \Tc when using MAGPIE+DSOAP features (c) and when using only MAGPIE features (d).}
\label{fig:totreldiff}
\end{figure}

Each entry in the \scicsd and the \scmp has the parameter \totreldiff which is a measure for how different the chemical formula of the SuperCon entry and the chemical formula of the original crystal structure before artificial doping were. The maximum allowed $\maxtotreldiff$ is an important cutoff parameter in the matching algorithm as explained in \autoref{sec:matching_algorithm}. Choosing an appropriate value of \maxtotreldiff is important because if it is chosen too small, not many SuperCon entries can be matched with crystal structures. In contrast, if it is chosen too big, the introduced bias due to artificial doping will be big because chemical formulas will be matched with crystal structures with which they are actually not related. 

\autoref{fig:totreldiff}a and b show a histogram of how many crystal structures in each dataset have a given \totreldiff. In \autoref{fig:totreldiff}a one can see that for the \scicsd the matched crystal structures decrease sharply with increasing \totreldiff. Extrapolating this curve to higher \totreldiff one can conclude that increasing the value of \maxtotreldiff would not have increased the number of matched crystal structures by much. This is less obvious in \autoref{fig:totreldiff}b for \scmp but there is a decreasing trend as well, and the number of data points with $\totreldiff \neq 0$ is smaller anyway. 

To study the influence of \totreldiff on the prediction accuracy we trained an XGB model on MAGPIE and MAGPIE+DSOAP features to compare the error of the test set dependent on the \totreldiff of each structure. \autoref{fig:totreldiff}c shows the distribution of the Symmetrical Mean Absolute Percentage Error (SMAPE) over the \totreldiff for a model trained on MAGPIE+DSOAP features. For comparison, the same plot is shown in \autoref{fig:totreldiff}d for a model trained only on MAGPIE features.

To analyze the influence of \totreldiff on the prediction accuracy one can look at \autoref{fig:totreldiff}c where the distribution of the Symmetrical Mean Absolute Percentage Error (SMAPE) was plotted over \totreldiff for each data point. We used the SMAPE here because it is a relative error which does not depend on the magnitude of \Tc. Thus the plotted distribution is free from the correlation of \totreldiff and \Tc. This is important because most cuprates have a \totreldiff between $0$ and $0.04$, therefore the MAE would have been very high around this \totreldiff without that the model would have actually been worse there, simply because cuprates tend to have a high \Tc. Intuitively one would expect that entries with a higher \totreldiff would have a higher error at the prediction, because these structures are only approximated with artificial doping. Such a correlation can not be seen in \autoref{fig:totreldiff}; the SMAPE seems to be independent of the \totreldiff. This shows that we chose \maxtotreldiff small enough so that artificial doping is a good approximation of the real crystal structures. Additionally \autoref{fig:totreldiff}d shows the same plot, but with an XGB model trained only on MAGPIE features instead of on MAGPIE+DSOAP features. This plot should definitely be independent of \totreldiff because the model was only trained on the chemical formula, which is not influenced by \totreldiff. The distribution looks very similar to the distribution of the model trained on MAGPIE+DSOAP features, which shows again that \totreldiff does not have a big influence on the dataset with structural features. We can also see this from the sorting criteria optimization (see \ref{sec:sorting_criteria_optimization}) because the \totreldiff did not seem to be an important sorting criteria there. 

In conclusion, the choice of \maxtotreldiff was sensible to get a lot of data points, while at the same time not decreasing the prediction accuracy. One could try to increase \maxtotreldiff until one notices a decrease in the performance, but probably this would not yield many more data points. 

\subsection{Normalized chemical formulas}
\label{sec:importance_of_normalization}

Normalizing the chemical formulas before matching is an important part of the matching algorithm by which a lot of SuperCon entries are matched which otherwise would not be matched. We will now analyze the influence of this normalization step. 

\begin{figure}
\centering
\includegraphics[width=\textwidth]{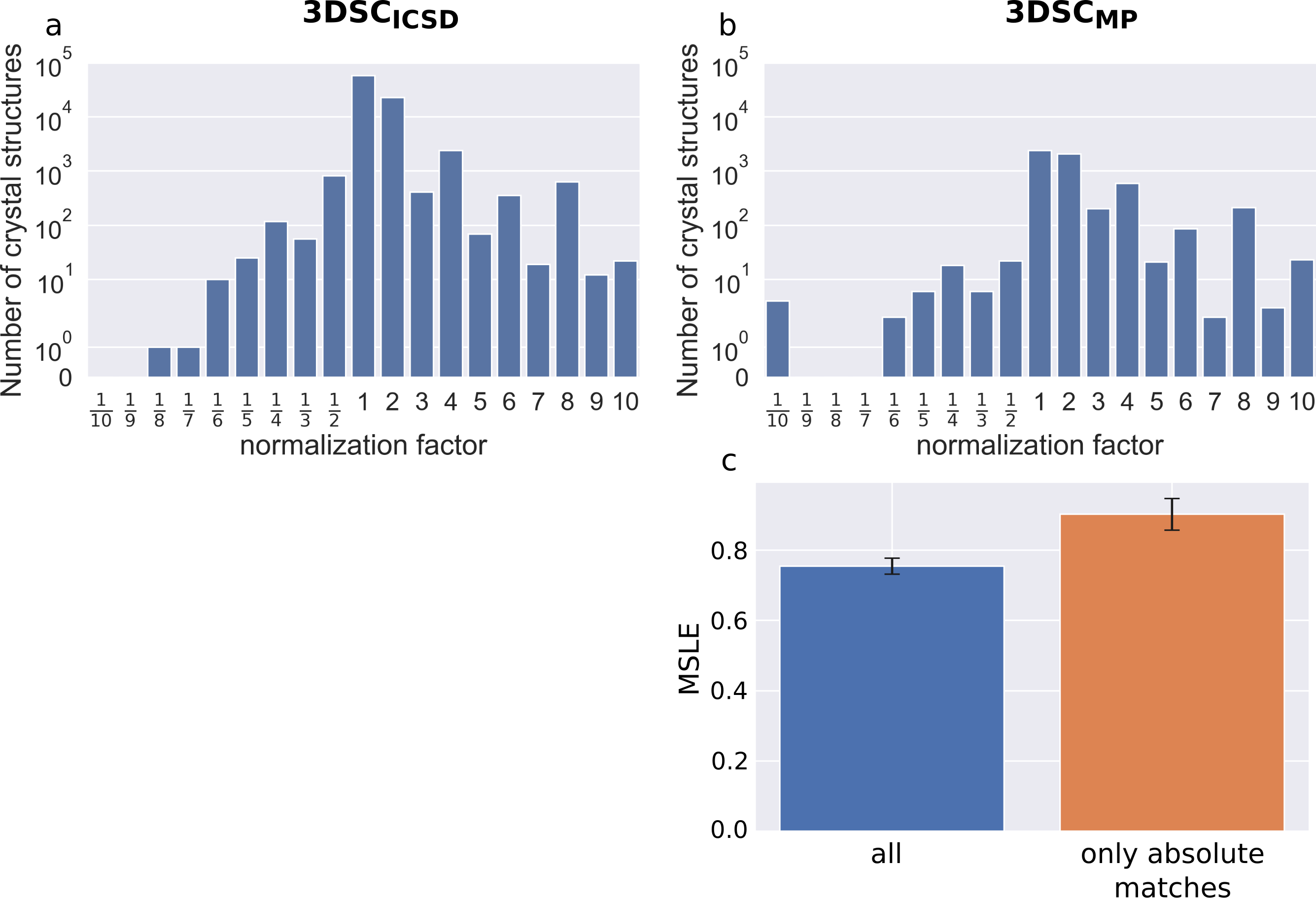}
\caption{Analysis of the importance of normalizing chemical formulas in the matching algorithm. The first row shows the normalization factor of chemical formulas of the SuperCon entry and the crystal structure for the \scicsd (a) and the \scmp (b). For the sake of clarity, the \x axis is shown only up to a factor of 10 in both directions. (c) A comparison of training and testing a model on all crystal structures vs only on crystal structures with absolute matches of the chemical formula for the \scmp. Shown are the mean and the standard error of the mean of 25 repetitions.}
\label{fig:abs_matches}
\end{figure}

\autoref{fig:abs_matches}a and b show the number of crystal structures with a given normalization factor between the chemical formula of the SuperCon entry and the chemical formula of the crystal structure for the \scicsd and the \scmp respectively as explained in \ref{si:normalizing_chemical_formulas}. These histograms show some more insight into matching normalized chemical formulas. One can see that in the histogram there are peaks with particularly many chemical formulas having a certain relative normalization factor. Besides the trivial peak at 1, there is a large peak at 2 and also at 4, 6 and 8. The peak at the 6 is a bit less pronounced. This behavior is consistent both for the ICSD and the Materials Project. These peaks are not symmetrical. That means there are a lot of cases where the chemical formula of the crystal structure is a multiple of $2^n$, but not the other way round. This is an indication that authors of SuperCon entries often did not know the exact primitive unit cell and simply wrote down the smallest integer chemical formula. 

We also trained an XGB model on the \scmp once with all data points and once only with data points where the chemical formula did not have to be normalized. This subset of data points decreases the number of matched SuperCon entries to  \SI{58}{\percent} (3358 SuperCon entries). The MSLE of this run is shown in \autoref{fig:abs_matches}c. The results show that training on the additional data points with normalized chemical formulas helps significantly in predicting \Tc. This shows that doing this normalization is overall beneficial, the additional data that is gained is worth the introduced bias.

We conclude that normalizing the chemical formulas seems to be overall beneficial. However, it is not a surprise that by having more data we get better results. For future studies it would be more interesting to also look at how the extrapolation performance changes for more difficult extrapolation settings, which is the actual benefit of training on crystal structures instead of only chemical formulas. 

\subsection{Random dropping of crystal structures and importance of symmetry features}
\label{si:random_dropping_and_fsym}

\begin{figure}
\centering
\includegraphics[width=\textwidth]{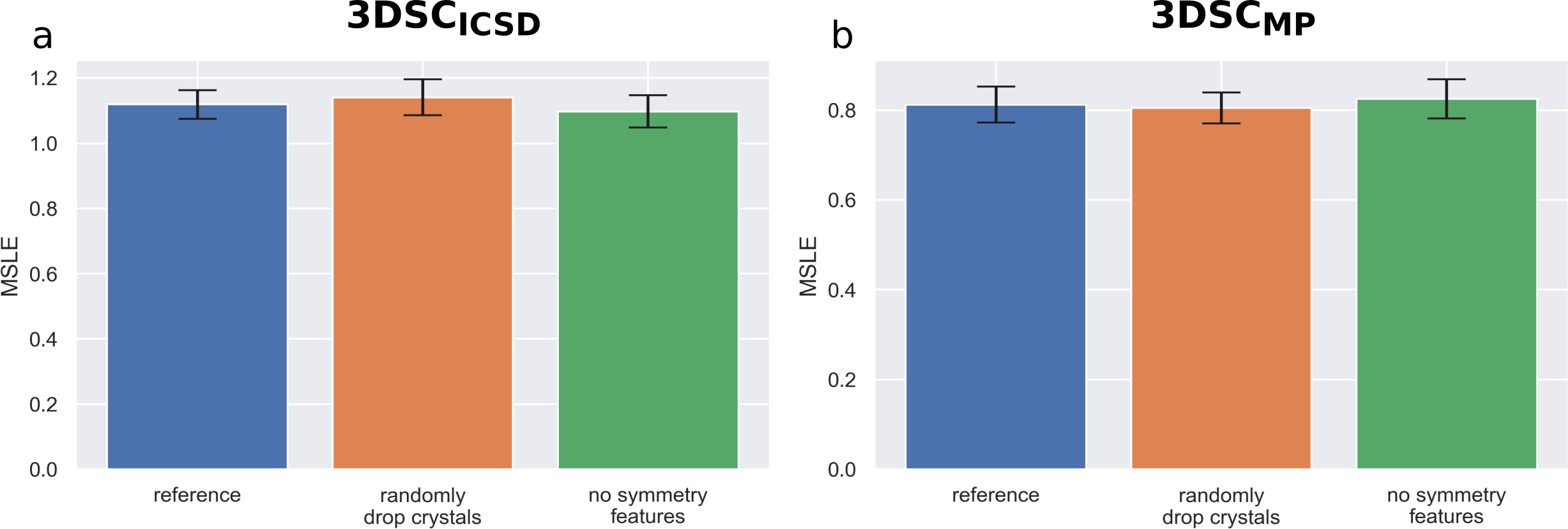}
\caption{Two independent ablation studies. a and b show a comparison of what happens if one either randomly drops all but one crystal structure for each SuperCon entry or leaves away the symmetry features \fsym for the \scicsd (a) and the \scmp (b). For comparison a reference run with all crystal structures and including symmetry features \fsym is shown.}
\label{fig:ablation_study}
\end{figure}

In this section we address two independent, little questions. First, we analyze the consequence of randomly dropping all but one crystal structure per SuperCon entry. This would be the simplest version of reducing the number of crystal structures per SuperCon entry to 1. Second, we analyze whether the symmetry features \fsym are informative.

We ran experiments with the XGB model as described above. The results in terms of the MSLE are shown in \autoref{fig:ablation_study}a and b. Each plot shows the reference run, the run with randomly dropped crystal structures and the run without symmetry features \fsym. 

Randomly dropping all but one crystal structure makes the MSLE in case of the \scicsd slightly worse, but not significantly. For the \scmp this has nearly no effect since there are only very few SuperCon entries with duplicate crystal structures. The fact that randomly dropping crystal structures does not significantly decrease performance is a good sign: On the one hand it allows for faster training without significantly losing performance. It also shows that most of the gain of information that one gets by including the crystal structure is already included if one has just one crystal structure per SuperCon entry. Also, randomly taking just one structure will probably on average be equal to diluting the few non-su\-per\-con\-duc\-ting structures with superconducting ones, because the dataset is very clustered with many close data points. On the other hand, it is likely that by randomly taking structures there will still be multiple cases where the non-su\-per\-con\-duc\-ting structure is chosen. Yet that means that there is still room for improvement if one would better choose the crystal structures, which should be studied in more detail in the future. 

Training without symmetry features \fsym seems to make the training for the \scicsd slightly better and for the \scmp slightly worse, but in both cases the changes are not statistically significant.

In conclusion, randomly dropping duplicate crystal structures does not significantly change the performance. The symmetry features on the other hand are not useful for the models with DSOAP features. 

\subsection{Sorting criteria optimization}
\label{sec:sorting_criteria_optimization}

In this section we will present how we developed the filtering criteria that were used to reduce the number of crystal structures per SuperCon entry as explained in \autoref{sec:matching_algorithm}. First, we chose 4 different criteria which might have an influence:
\begin{enumerate}

\item \totreldiff: Entries with a lower \totreldiff are potentially less biased by the artificial doping. This sorting criteria is semi-continuous: Technically it is a continuous number, but due to the integer number of atoms in most chemical formulas, there will still often be multiple crystal structures of one SuperCon entry with the same \totreldiff.
\item Whether or not the chemical formula had to be normalized in order to match: Crystal structures and SuperCon entries potentially match better if their chemical formulas do not have to be normalized to match. This is a binary sorting criteria which means that after sorting by this criteria there will usually still be many crystal structures with the same ranking.
\item Energy above the hull \ehull (Materials Project): This criteria applies only to the \scmp because only the Materials Project has \ehull given. \ehull is a property that can be used to predict the stability of the phase of a crystal structure. It is a continuous number and it happens only very rarely that two crystal structures of the same SuperCon entry have the same \ehull. Therefore, after sorting and filtering by \ehull, there will nearly never be some duplicate crystal structures left. 
\item \explicittcry (ICSD): This criterion only applies to the \scicsd. As explained in \autoref{sec:data} not all of the ICSD crystal structures had \Tcry explicitly given. We assumed that crystal structures which did have \Tcry explicitly given are more trustworthy. This sorting criteria is categorical.
\item One additional option was to use all data points without any filtering.
\end{enumerate}

\begin{figure}
\centering
\includegraphics[width=\textwidth]{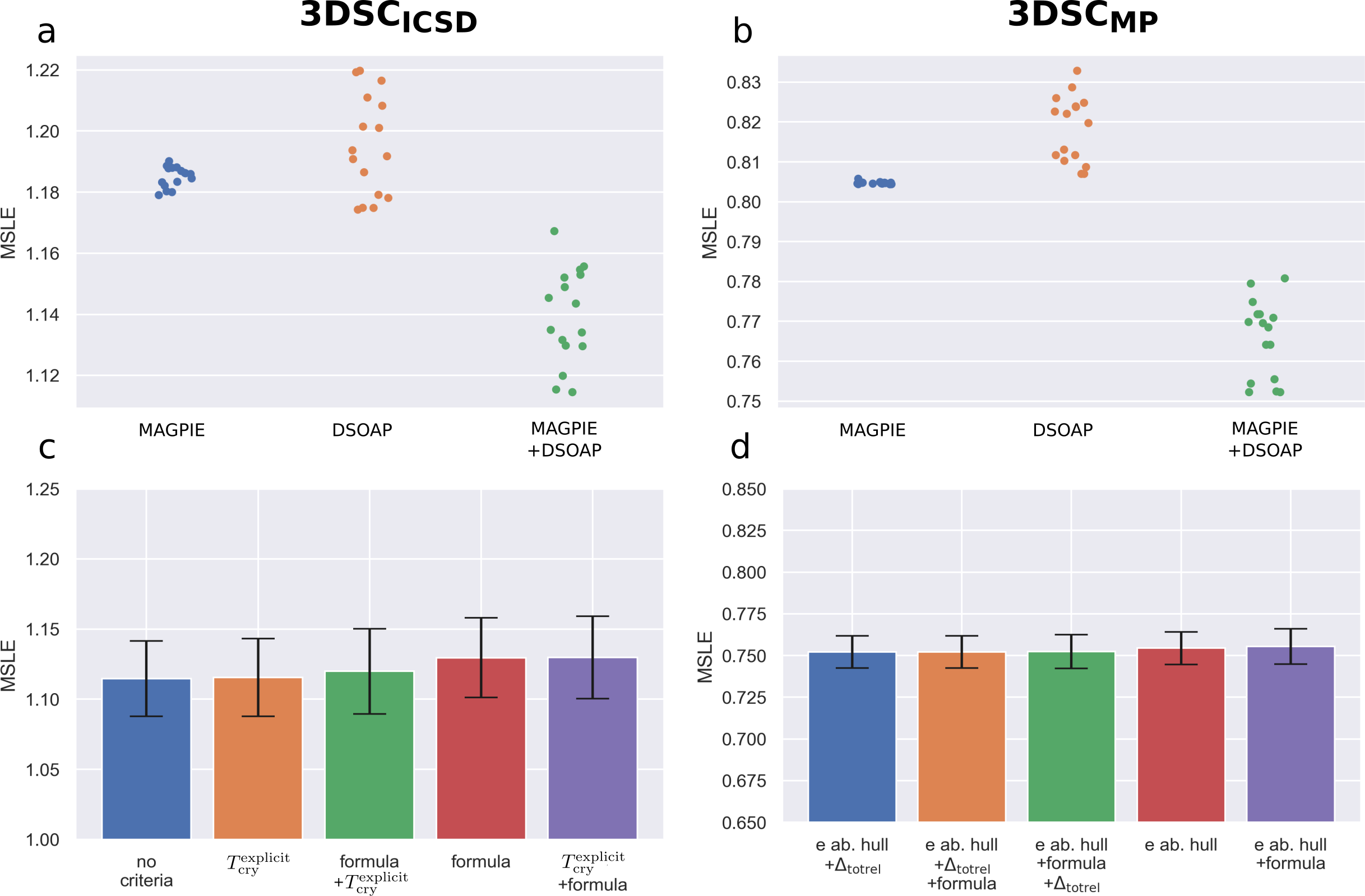}
\caption{Results of the sorting criteria optimization. (a) and (b) show the MSLE of all runs of the sorting criteria optimization with different sorting criteria for the \scicsd (a) and the \scmp (b). Each data point is the mean of 25 repetitions for the \scicsd and 100 repetitions for the \scmp. (c) and (d) show mean and error of the mean of the top 5 sorting criteria for the \scicsd (c) and the \scmp (d) for the run with MAGPIE+DSOAP features sorted by their MSLE.}
\label{fig:3dsc_sorting_criteria}
\end{figure}

The results of the sorting criteria hyperparameter optimization are shown in \autoref{fig:3dsc_sorting_criteria}a to d. Figures a and b show for each run the mean of all 100 and 25 cross validation repetitions for the \scicsd and \scmp respectively. We plotted all runs of the grid search so that one can be certain that a particularly good result is not only due to overfitting to the test set. 

From the plots in \autoref{fig:3dsc_sorting_criteria}a and b one can see that the MAGPIE+DSOAP features are consistently better than only MAGPIE or only DSOAP features, for both datasets. This shows that including structural information indeed helps the model in predicting the critical temperature \Tc. Interestingly, using only DSOAP features often leads to worse results than using only MAGPIE features. One possible reason is that MAGPIE features include information about chemical closeness of elements, which helps predicting rare elements. As shown in \autoref{fig:stats_dataset} (e) and (f) there are a lot of these rare elements in the dataset which do not appear very often.

We also tried out to incorporate the electronic features of the \scmp, but this did not significantly improve the results so we did not include them in the analysis in the main part. The electronic features that we tried were the band gap, energy, energy per atom, formation energy per atom, total magnetization, number of unique magnetic sites and the true total magnetization as recorded in the Materials Project. Note that we did not try to use the Fermi energy \efermi because it was not given for all crystal structures in the \scmp.

It is noticeable that for the \scicsd in \autoref{fig:3dsc_sorting_criteria}a the variance of the runs with MAGPIE features is greater than for the \scmp in  \autoref{fig:3dsc_sorting_criteria}b. In theory, each run which uses only MAGPIE features should have exactly the same performance because we controlled that every split has exactly the same SuperCon entries and only the used crystal structures are different between the runs. However, because each SuperCon entry can have a different number of crystal structures, it also appeared a different number of times for the model. In theory this should not matter because we were passing a sample weight with each crystal structure to weigh each SuperCon entry the same. However, we suspect that this randomness is due to the data bagging in the XGB algorithm. Due to the data bagging, different partitions of the data will be used for each decision tree if the dataset is not exactly the same. This effect is much stronger for the \scicsd than for the \scmp because the \scicsd has much more crystal structures per SuperCon entry than the \scmp. One could probably mitigate this issue by fixing that all crystal structures of one SuperCon entry will all be given to the same decision tree.

\autoref{fig:3dsc_sorting_criteria}c and d show the top five sorting criteria ordered by their MSLE with mean and error of the mean for the \scicsd and \scmp respectively. In the \scicsd (c) simply using all data points indeed is the best option. The second best option with no significant difference in performance is sorting by \explicittcry. Because the performance of the two runs has no significant difference we decided to use the latter, because this reduced the number of crystal structures in the \scicsd database from approximately 140,000 to approximately 80,000 and makes consecutive training faster and less memory intensive. 

For the \scmp (d) it seems that sorting by the energy above the hull \ehull is the by far most important criteria. Even though the following option also have other criteria after \ehull, these criteria effectively do not matter because \ehull is a continuous float value. It is interesting that the algorithm clearly chooses structures with low \ehull to be more informative in this dataset. The Materials Project contains a large number of theoretical structures and the Materials Project website marks experimentally confirmed structures, but this parameter is not accessible in the API. The structure with the minimum \ehull is usually experimentally confirmed and the most usual structure for this material, so it might be that the algorithm just focused on excluding overly theoretical structures. Finding out the exact role of \ehull in this optimization would be an interesting aspect of further research. 

In conclusion, we chose the criteria \explicittcry as the sorting criteria to use for the \scicsd. This matches the 9,150 SuperCon entries in this dataset with 86,490 crystal structures. For the \scmp we chose the criteria of sorting first by \ehull and then by \totreldiff as the sorting criteria for this dataset. This matches the 5,759 SuperCon entries in this dataset with 5,773 crystal structures. 

%\begin{figure}[ht]
%\centering
%\includegraphics[width=\textwidth]{stream}
%legend text.}
%\label{fig:stream}
%\end{figure}

\end{document}